\documentclass[journal=apchd5,manuscript=article]{achemso}
\usepackage[version=3]{mhchem} % Formula subscripts using \ce{}
\usepackage{amssymb,epsfig,setspace,color}

\graphicspath{{figs/}}

\newcommand{\be}{\begin{equation}}
\newcommand{\ee}{\end{equation}}
\newcommand{\bg}{\begin{equation}}
\newcommand{\eg}{\end{equation}}
\newcommand{\bdm}{\begin{displaymath}}
\newcommand{\edm}{\end{displaymath}}
\newcommand{\bea}{\begin{eqnarray}}
\newcommand{\eea}{\end{eqnarray}}
\newcommand{\beas}{\begin{eqnarray*}}
\newcommand{\eeas}{\end{eqnarray*}}
\newcommand{\ba}{\begin{array}}
\newcommand{\ea}{\end{array}}

\newcommand{\bfg}{\begin{figure}}
\newcommand{\efg}{\end{figure}}

\newcommand{\mb}{\mbox}

\newtheorem{lm}{Lemma}
\newtheorem{cl}{Corollary}
\newtheorem{df}{Definition}

\newcommand{\blm}{\begin{lm}}
\newcommand{\elm}{\end{lm}}
\newcommand{\bcl}{\begin{cl}}
\newcommand{\ecl}{\end{cl}}
\newcommand{\bdf}{\begin{df}}
\newcommand{\edf}{\end{df}}
\newcommand{\brk}{\begin{rm}}
\newcommand{\erk}{\end{rm}}
\newcommand{\lb}{\label}

\newcommand{\al}{\alpha}

\newcommand{\veps}{\varepsilon}

\newcommand{\ld}{\lambda}

\newcommand{\vE}{{\bf E}}

\author{Song Sun}
\affiliation{Microsystem and Terahertz Research Center, China Academy of Engineering Physics, No. 596, Yinhe Road, Shuangliu, Chengdu, 610200, China}
\alsoaffiliation{Institute of Electronic Engineering, China Academy of Engineering Physics, Mianyang, 621999, China}
\altaffiliation{Contributed equally to this work}
\email{sunsong@mtrc.ac.cn}

\author{Ilia L. Rasskazov}
\affiliation{The Institute of Optics, University of Rochester, Rochester, NY 14627, USA}
\altaffiliation{Contributed equally to this work}
\email{irasskaz@ur.rochester.edu}

\author{P. Scott Carney}
\affiliation{The Institute of Optics, University of Rochester, Rochester, NY 14627, USA}

\author{Taiping Zhang}
\affiliation{Microsystem and Terahertz Research Center, China Academy of Engineering Physics, No. 596, Yinhe Road, Shuangliu, Chengdu, 610200, China}
\alsoaffiliation{Institute of Electronic Engineering, China Academy of Engineering Physics, Mianyang, 621999, China}

\author{Alexander Moroz}
\affiliation{Wave-scattering.com}
\email{wavescattering@yahoo.com}

\title{The critical role of shell in enhanced fluorescence of metal-dielectric core-shell nanoparticles}

\begin{document}

\begin{tocentry}
\includegraphics{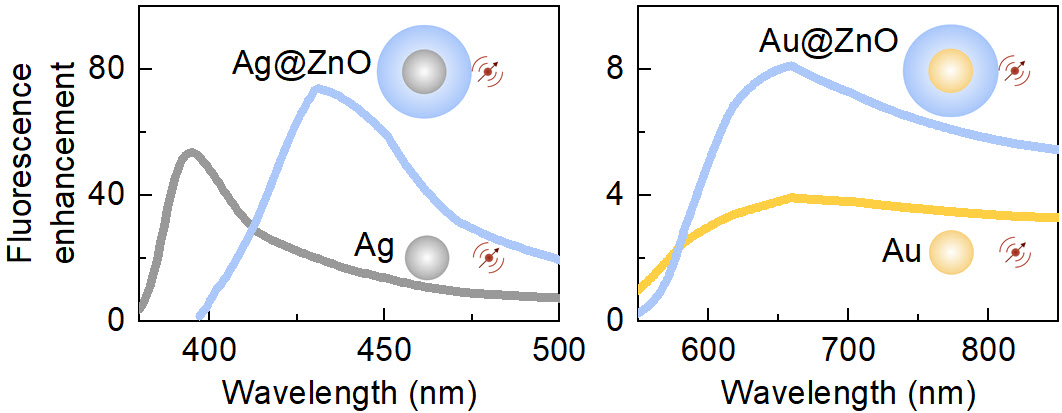}
\end{tocentry}

\begin{abstract}
Large scale simulations are performed by means of the transfer-matrix method to reveal optimal conditions for metal-dielectric core-shell particles to induce the largest fluorescence on their surfaces. 
With commonly used plasmonic cores (Au and Ag) and dielectric shells (SiO${}_2$, Al${}_2$O${}_3$, ZnO), optimal core and shell radii are determined to reach maximum fluorescence enhancement for each wavelength within $550$--$850$~nm (Au core) and $390$--$500$~nm (Ag core) bands, in both air and aqueous hosts. 
The peak value of the maximum achievable fluorescence enhancement factors of core-shell nanoparticles, taken over entire wavelength interval, increases with the shell refractive index and can reach values up to 9 and 70 for Au and Ag cores, within $600-700$~nm and $400-450$~nm wavelength ranges, respectively, which is much larger than that for corresponding homogeneous metal nanoparticles. 
Replacing air by an aqueous host has a dramatic effect of nearly halving the sizes of optimal core-shell configurations at the peak value of the maximum achievable fluorescence. 
In the case of Au cores, the fluorescence enhancements for wavelengths within the first near-infrared biological window (NIR-I) between 700 and 900~nm can be improved twofold compared to homogeneous Au particle when the shell refractive index $n_s\gtrsim 2$. 
As a rule of thumb, the wavelength region of optimal fluorescence (maximal nonradiative decay) turns out to be red-shifted (blue-shifted) by as much as $50$~nm relative to the localized surface plasmon resonance wavelength of corresponding optimized core-shell particle. Our results provide important design rules and general guidelines for enabling versatile platforms for imaging, light source, and biological applications.
\end{abstract}

%%%%%%%%%%%%%%%%%%%%%%%%%%
\section{Introduction}
%%%%%%%%%%%%%%%%%%%%%%%%%%

Fluorescence-based spectroscopy and imaging techniques have become a promising solution to meet the demands of various emerging applications such as single molecule detection~\cite{Stehr2019,Ray2018}, early diagnosis~\cite{Bower2018,Garcia2018,Park2019a}, food and drug safety~\cite{Andersen2008}. 
The advantage of fluorescence emitters to label the target species at a molecular level makes it an ideal tool for fingerprint tags~\cite{Lu2018a}. 
Fluorescence spontaneous emission also serves as the foundation of advanced light sources such as micro/nano light emitting diodes (LEDs) for high-resolution displays~\cite{Schmidt2017,Yang2015b} and single photon sources for quantum photonics~\cite{Reimer2019,Xu2019}. 
Independent spatial and temporal radiation characteristics of fluorescence emitters have been employed in a super-resolution imaging.
Despite all those attractive features, intrinsic fluorescence emission is very weak, and challenges the development of high performance devices. 

Plasmonic nanostructures hold a great potential in enhancing fluorescence emission~\cite{Anger2006,Bharadwaj2007,Bharadwaj2009,Fothergill2018,Li2017f}. 
On one hand, collective electron oscillations on the surface of plasmonic nanostructure can generate a strong local electric field enhancement to boost the excitation rate of the fluorescence emitter~\cite{Ford1984,Anger2006,Bharadwaj2007,Sun2016d,Dong2015}. 
On the other hand, the presence of metal nanostructure in the vicinity of the fluorescence emitter affects the local density of optical states (LDOS), thereby tailoring the radiative and nonradiative decay rates~\cite{Ruppin1982,Ford1984,Chew1987,Chew1988,Kim1988,Moroz2005,Moroz2005a,Girard2010,Guo2016b}. 
An optimal fluorescence enhancement factor requires a delicate balance of the excitation, radiative and nonradiative decay rates~\cite{Anger2006,Bharadwaj2007,Sun2017b,Ringler2008}. 
To date, plasmonic structures have been developed to obtain fluorescence enhancement, also termed metal-enhanced fluorescence~\cite{Geddes2002}, such as metallic layered structures~\cite{Akimov2017,Huang2011}, waveguides~\cite{Wu2019b}, ordered structures~\cite{Brolo2005,Guo2008}, nanoantennas~\cite{Ma2018,Kinkhabwala2009,Sun2018,Sun2019a}, nanoparticles~\cite{Anger2006,Bharadwaj2007,Tam2007,Fan2016} to name just a few.  

Compared to other plasmonic alternatives, core-shell nanoparticles possess unique advantages owing to their mass production capability with low cost chemical synthesis methods~\cite{Oldenburg1998,Graf2002,Graf2003,Tovmachenko2006,Aslan2007,Guerrero2011,Montano-Priede2017a,Xu2018}. 
With regard to tuning the localized surface plasmon resonance (LSPR) wavelength, metal shell particles are superior to dielectric shell particles. 
As it can be qualitatively understood already from the quasi-static analysis,~\cite{Neeves1989} in the metal shell case one can tune the dipole LSPR between 
Re $\veps_s=-3(\veps_c+2\veps_h)/\{2[1-(r_c/r_s)^3]\}$ for $r_s\to r_c$ and Re $\veps_s=-2\veps_h$ for $r_c\to 0$,
where $r_c$ and $r_s$ are core and shell radii, and $\veps_c$, $\veps_s$ and $\veps_h$ correspond to dielectric permittivities of a core, shell and host medium, respectively. For Ag and Au shells, this translates into the whole visible and near-infrared range simply by a control of the core-shell morphology, i.e. of the ratio $r_c/r_s$~\cite{Neeves1989,Oldenburg1998,Averitt1997,Graf2002}. In the dielectric shell case, the tunability limits are set between Re $\veps_c=-2\veps_h$ for $r_s\to r_c$ and Re $\veps_c=-2\veps_s$ for $r_c\to 0$,~\cite{Neeves1989} and the resulting tunability is narrower and typically of the order of $\sim 100$~nm. For the same reason, initial search of fluorescence enhancement focused more on metal shell particles, whereas the use of metal-dielectric core-shell particles, apart of some preliminary work~\cite{Tovmachenko2006,Aslan2007,Reineck2013,Lu2014a,Wang2014a,Pang2015,Planas2016,Walters2018,Niu2018,Camacho2016,Meng2018}, has still remained to be underestimated in the current literature. 

In what follows, we show that the neglect of dielectric shell particles has been largely undeserved.
We searched for experimentally feasible metal-dielectric core-shell configurations with common Au and Ag cores and widely available dielectric shell materials (SiO${}_2$, Al${}_2$O${}_3$, ZnO) whose refractive indices are higher than that of the host medium (air or water), for optimal fluorescence enhancements. 
The outcome is that, in spite of relatively narrow tunability of the LSPR wavelength of those particles, there is still enough of design flexibility left for optimally designed nanoparticles to enable (i) comparable or even larger fluorescence enhancement as metal shells~\cite{Arruda2017a} and (ii) significantly enhanced fluorescence compared to homogeneous metal particles~\cite{Tovmachenko2006,Aslan2007,Camacho2016,Niu2018,Meng2018}, due to the efficient tailoring of the electric near-field and fluorescence decay rates by dielectric shell.
Furthermore, the dielectric shell of a metal-dielectric core-shell nanoparticle (also called shell-isolated nanoparticle~\cite{Meng2018}) provides a convenient way to separate the fluorescence emitter and the metallic component with a predetermined distance, thus avoiding the quenching problem~\cite{Tovmachenko2006,Anger2006,Noginov2009,Acuna2012,Reineck2013,Walters2018}.
In our simulations, we have employed a rigorous and computationally fast transfer matrix method, which can be viewed as an extension of the theory for homogeneous particles \cite{Ruppin1982,Chew1987,Chew1988,Kim1988} to a multilayered case with an arbitrary number of layers, to obtain the radiative and nonradiative decay rates~\cite{Moroz2005,Moroz2005a}, the electric field distribution~\cite{Rasskazov19JOSAA}, and, as a result, the fluorescence enhancement factor.

%%%%%%%%%%%%%%%%%%%%%%%%%%
\section{Methods}
\begin{figure}
    \centering
    \includegraphics{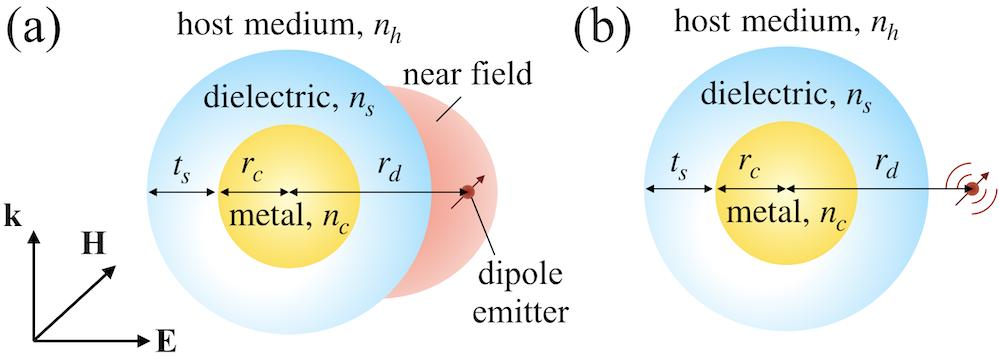}
    \caption{
    A conventional two-step model for nanoparticle enhanced fluorescence: (a) excitation process under plane wave illumination, and (b) emission process with dipole radiation. 
    The origin of coordinates is located at the particle center.}
    \label{fig:scheme}
\end{figure}

\lb{sc:meth}
%%%%%%%%%%%%%%%%%%%%%%%%%%
Core-shell nanoparticle enhanced fluorescence can be described in two steps as shown in Figure~\ref{fig:scheme}, with the fluorescence emitter being modelled as an oscillating dipole~\cite{Ruppin1982,Ford1984,Chew1987,Chew1988,Kim1988,Moroz2005}.
First, the core-shell nanoparticle locally enhances an electric field under a plane wave excitation, thereby amplifying the excitation rate of the fluorescence emitter in its proximity. 
Second, after being excited, the emitter itself radiates, and mutual interaction with the nanoparticle modifies the radiative and nonradiative decay rates of an emitter~\cite{Chew1988,Bharadwaj2007,Sun2016d,Sun2019a}. 
Note that the fluorescence excitation and emission processes are treated independently (i.e. weak coupling) and the emitter is assumed to be below saturation~\cite{Anger2006,Bharadwaj2007}.

%%%%%%%%%%%%%%
%%%%%%%%%%%%%%

\subsection{Excitation process: electric field enhancement}
%%%%%%%%%%%%%%%%%%%%%%%%%%
Under a plane wave illumination, the electric field in $j$-th layer of a general multilayered core-shell nanoparticle, from the core ($j=1$) up to the host medium ($j=N+1$), where the number of layers in our case is $N=2$, is expanded into multipole expansion~\cite{Moroz2005}:

\begin{equation}
\label{eq:fld}
{\bf E}_p ({\bf r}) = \sum_{p,L} \left[ A_{p L} (j) {\bf J}_{p L}(k_j, {\bf r}) + B_{p L}(j) {\bf H}_{p L}(k_j, {\bf r})\right] \ .
\end{equation}
Here  
$L = \ell,m$ is the composite angular momentum index with $\ell$ and $m$ being the usual orbital and magnetic angular momentum numbers, the respective
$p=E$ and $p=M$ denote electric (or TM) and magnetic (or TE) polarizations,
$k_j = 2\pi n_j / \lambda$ is the wavenumber in $j$-th shell with refractive index $n_j$,
$\lambda$ is the incident wavelength,
$A_{p L} (j)$ and $B_{p L} (j)$ are the expansion coefficients, 
${\bf J}_{p L}(k_j, {\bf r})$ and ${\bf H}_{p L}(k_j, {\bf r})$ are vector multipoles~\cite{Moroz2005,Rasskazov19JOSAA}.
The expansion coefficients $A_{p L} (j)$ and $B_{p L} (j)$ can be obtained by matching the tangential components of the electromagnetic fields at each interface by implementing the recursive algorithm.

\subsection{Emission process: radiative and nonradiative decay rates}
%%%%%%%%%%%%%%%%%%%%%%%%%%

The interaction between the dipole emission and the core-shell nanoparticle can be analytically solved using the transfer matrix method~\cite{Moroz2005,Rasskazov18OMEx}.
A dipole emitter is assumed to be located at the radial distance $r_d$, which can be inside the shell or outside the core-shell nanoparticle~\cite{Tovmachenko2006,Aslan2007,Noginov2009,Guerrero2011,Xu2018,Lu2014a,Pang2015,Planas2016,Wang2014a}. 
For non-magnetic core-shell particle and host, the radiative, $\gamma_{\rm rad}$, and nonradiative decay rates, $\gamma_{\rm nrad}$, normalized with respect to the radiative decay rate in the free space (assumed to have the host permittivity), $\gamma_{{\rm rad};0}$, can be obtained as~\cite{Moroz2005}:

\begin{equation}
    \begin{split}
        \dfrac{\gamma_{\rm rad}^{\perp}}{\gamma_{{\rm rad};0}} & = \dfrac{3}{2x^4_d} \dfrac{n_d}{n_h}
        \sum_\ell \ell \left( \ell+1 \right) \left(2\ell+1\right) \left| f_{E\ell}(x_d) \right|^2 \ , \\
        \dfrac{\gamma_{\rm rad}^{\parallel}}{\gamma_{{\rm rad};0}} & = \dfrac{3}{4x^2_d} \dfrac{n_d}{n_h} 
        \sum_\ell \left(2\ell+1\right)\left[ \left|f_{M\ell}(x_d) \right|^2 + \left| f'_{E\ell}(x_d) \right|^2 \right] \ , \\
        \frac{\gamma_{\rm nrad}^{\perp}}{\gamma_{{\rm rad};0}} & = \frac{3k_d^3}{2x^4_d}\dfrac{1}{n_d n_h} {\rm Im} \left(n_a^2\right) \sum_\ell \ell \left(\ell+1\right) \left(2\ell+1\right) I_{E\ell} \left| d_{E\ell}(x_d) \right|^2 \ , \\
        \frac{\gamma_{\rm nrad}^{\parallel}}{\gamma_{{\rm rad};0}} & = \frac{3k_d^3}{4x^2_d}\dfrac{1}{n_d n_h}{\rm Im}\left(n_a^2\right) \sum_\ell \left(2\ell+1\right) \left[ I_{M\ell} \left| d_{M\ell}(x_d) \right|^2 + I_{E\ell} \left| d'_{E\ell}(x_d) \right|^2 \right] \ ,
    \end{split}
    \label{eq:decay}
\end{equation}
where the respective ``$\perp$'' and ``$\parallel$'' indicate the perpendicular (radial) and parallel (tangential) dipole orientation relative to the particle surface, 
$n_a$ is the refractive index of the dissipative component with a non-zero imaginary part (i.e. the plasmonic metal core in our case),
$n_h$ is the refractive index of the host medium,
$f_{p\ell}(x_d)$ and $d_{p\ell}(x_d)$ are linear combinations of Riccati-Bessel functions, 
$x_d = k_d r_d$, and $k_d = 2\pi n_d /\ld$, where the subscript $d$ corresponds to location of a dipole emitter.
The symbols $I_{p\ell}$ represent volume integrals taken over any absorbing region, i.e. in our case over the entire metal core.
Finally, the prime denotes the differentiation with respect to the argument in parentheses.

\subsection{Fluorescence enhancement}
%%%%%%%%%%%%%%%%%%%%%%%%%%

For light intensities below dye saturation, the fluorescence enhancement factor can be expressed as the product of the excitation rate and the quantum yield, which is essentially related to the electric field distribution and dipole decay rates of the core-shell nanoparticle:

\begin{equation}
    \dfrac{\eta_{\rm em}}{\eta_{{\rm em};0}} = \dfrac{\gamma_{\rm exc}}{\gamma_{\rm exc;0}} \dfrac{q}{q_0} \cdot
    \label{eq:bff}
\end{equation}
Here the subscript ``0'' indicates the respective quantity in the free space. 
The excitation rate can be expressed as $\gamma_{\rm exc} \propto |{\bf E} \cdot {\bf d} |^2$, where ${\bf E}$ is the local electric field at the emitter’s position which can be obtained from eq~\ref{eq:fld}, and ${\bf d}$ represents the dipole moment of the emitter. 
In our simulations we shall use both orientationally averaged electric field intensity and orientationally averaged decay rates.
The orientationally averaged electric field intensity is to be understood as the field intensity averaged over a spherical surface of a given fixed radius $r$~\cite{Rasskazov19JOSAA}, whereas an orientationally averaged decay rate is determined at a fixed dipole position by averaging over all possible dipole orientations. The surface integrals of intensities can be performed analytically~\cite{Rasskazov19JOSAA}, whereas orientation-averaged decay rates can be determined directly from eq~\ref{eq:decay} as $\gamma_{\rm nrad;rad} = ( \gamma^{\perp}_{\rm nrad;rad} + 2\gamma^{\parallel}_{\rm nrad;rad})/3$. Fluorescence enhancement at a given radial position will be determined by substituting into eq~\ref{eq:bff} an average quantum yield
\begin{equation}
    q = \dfrac{\gamma_{\rm rad}/\gamma_{{\rm rad};0}}{\gamma_{\rm rad}/\gamma_{{\rm rad};0} + \gamma_{\rm nrad}/\gamma_{{\rm rad};0} + (1-q_0)/q_0} \ ,
    \label{eq:QY}
\end{equation}
with the intrinsic quantum yield $q_0$ assumed, for simplicity, to be unity. The quantum yield  accounts for the competition between the radiative and nonradiative decay rates. Note that $q\ne \left( q^{\perp} + 2q^{\parallel}\right)/3$.

The above makes it clear that, although the core-shell far-field (i.e. scattering) properties can be, at least qualitatively, understood by the quasi-static analysis~\cite{Neeves1989,Chung2009}, the influence of near-fields on fluorescence is much more involved. 
Below, a systematic investigation is conducted taking into account these near-field effects to find optimal conditions for fluorescence enhancement.

\section{Optimized core-shell configurations for fluorescence enhancement}
\lb{sc:optim}
%%%%%%%%%%%%%%%%%%%%%%%%%%

The speed and robustness of our method allows us to perform an optimization study scanning over up to $\approx 10^5$ different core-shell configurations for each wavelength and for each shell material.
The schematic of the structure is shown in Figure~\ref{fig:scheme} displaying the core-shell nanoparticle with a metal core (with radius $r_c$ and refractive index $n_c$) surrounded by a dielectric shell (with thickness $t_s$ and refractive index $n_s$).
The nanoparticle is embedded in a homogeneous medium with a refractive index $n_h$, which is set to be air ($n_h=1$) or water ($n_h=1.33$) in what follows.
Au and Ag were selected as the core materials. 
For the sake of comparison with earlier results, Palik et al.~\cite{Palik1998} data of Au and Ag dielectric function were used in our simulations. 
Noteworthy, optimally synthesized Au and Ag cores may exhibit lower losses~\cite{McPeak2015}, thereby facilitating even higher fluorescence enhancements.
%%%%%%%%%%%%%%%%%%%%%%%%%%
\begin{figure}[t]
    %\centering
    \includegraphics{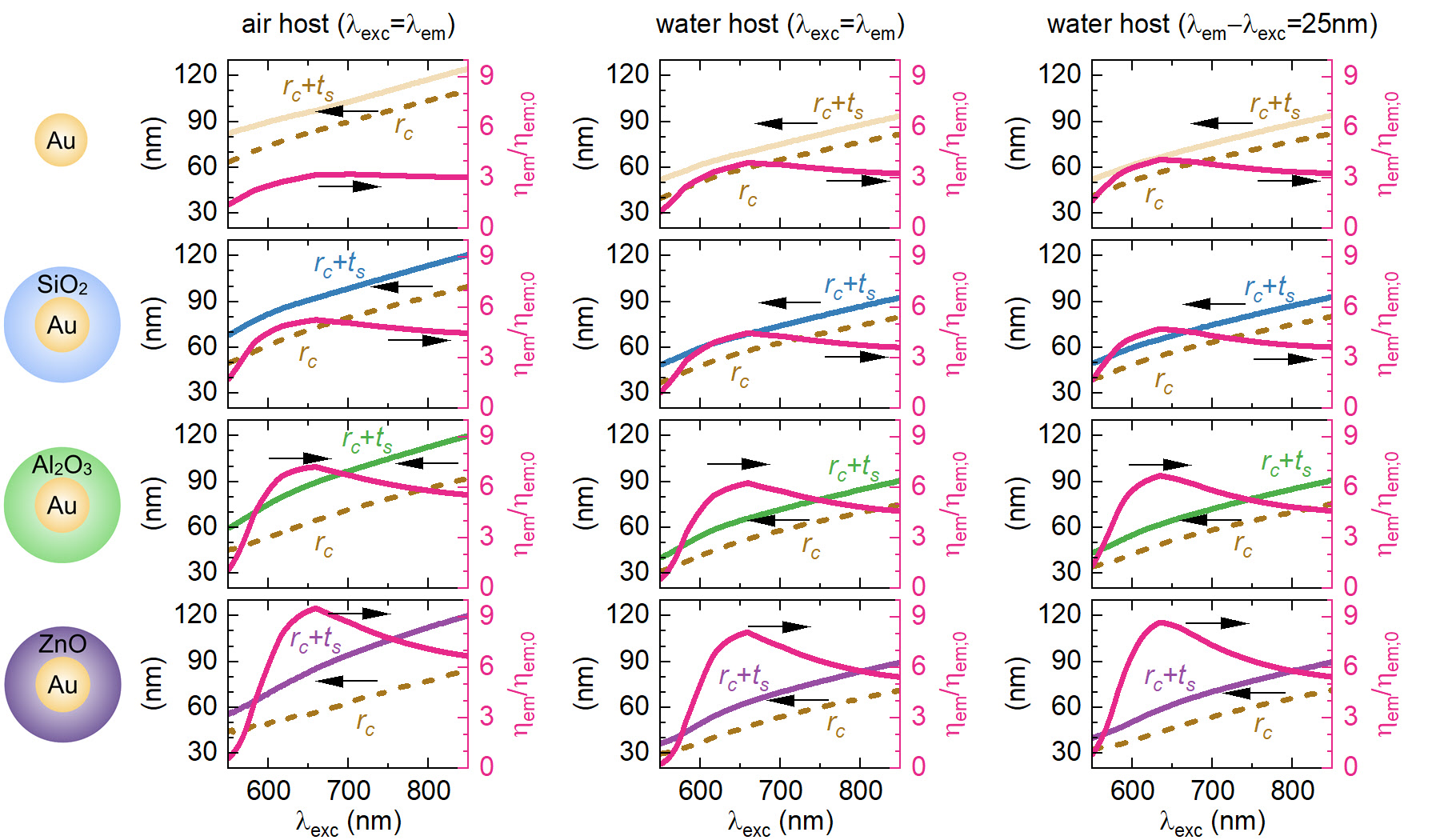}
    \caption{
    Optimal parameters ($r_c$,$t_s$) of Au@dielectric core-shell nanoparticles having maximum achievable fluorescence enhancement at a given excitation wavelength $\ld_{\rm exc}$ for a dipole emitter located at $0.75$~nm distance (given that a dye size is typically $1-2$~nm) from a surface of a shell for nanoparticles with different materials of a shell, from top to bottom: 
    bare Au nanoparticle, 
    SiO${}_2$ shell, 
    Al${}_2$O${}_3$ shell, and 
    ZnO shell.
    Nanoparticles are embedded in air or water host medium, and dyes are assumed to have zero or 25~nm Stokes shift ($\ld_{\rm em} - \ld_{\rm exc}$, where $\ld _{\rm em}$ is the emission wavelength).
    For the bare Au nanoparticle, the actual distance between the dye and a metal surface is $(t_s + 0.75)$~nm, which allows for a relevant comparison with core-shells. 
    Both orientationally averaged electric field intensity~\cite{Rasskazov19JOSAA} and average dipole orientation are used at each wavelength.}
    \label{fig:optAu}
\end{figure}
%%%%%%%%%%%%%%%%%%%%%%%%%%

\begin{figure}[t]
    \centering
    \includegraphics{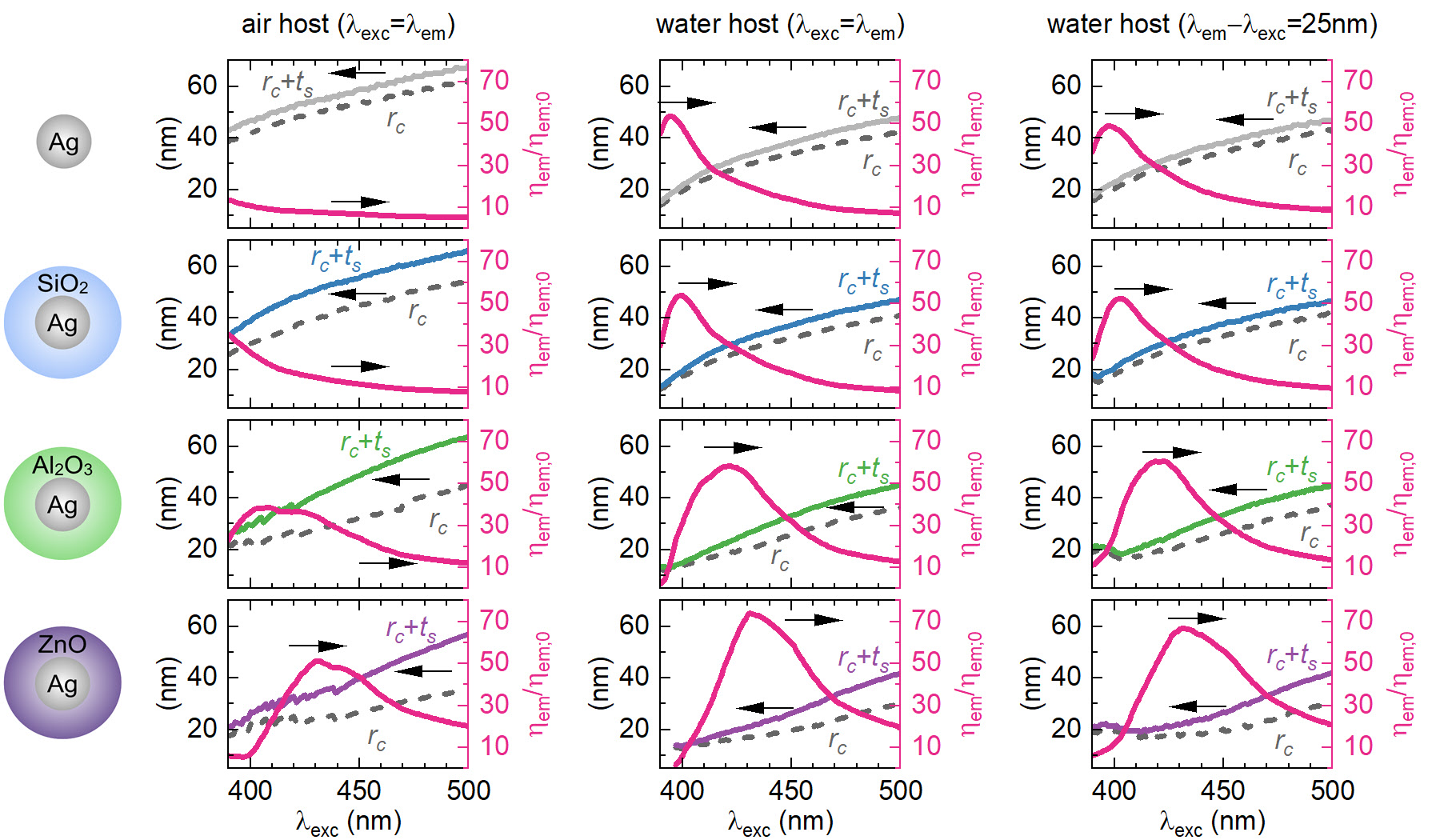}
    \caption{
    Same as in Figure~\ref{fig:optAu}, but for Ag as the core material. 
    Due to strong absorption in Ag close to the LSPR of homogeneous Ag sphere, one observes pronounced fluorescence quenching, $\eta_{\rm em}/\eta_{\rm em;0} \to 0$, for $\ld_{\rm exc}<400$~nm in aqueous host. 
    Almost an order of magnitude higher maximum fluorescence enhancement ($\approx 70$) can be achieved for Ag@dielectric core-shells than for Au@dielectric counterparts ($\approx 9$) of Figure~\ref{fig:optAu}.}
    \label{fig:optAg}
\end{figure}

Optimization results are obtained by scanning over different Au (Figure~\ref{fig:optAu}) and Ag (Figure~\ref{fig:optAg}) core radii within the interval $20 \le r_c\le 130$~nm and $5 \le r_c\le 70$~nm, correspondingly. 
We allow the shell thickness to vary within the interval $0 < t_s\le 40$~nm and perform optimization for three different shell refractive indices $1.45$, $1.76$, $2.00$, corresponding to typical values of SiO${}_2$, Al${}_2$O${}_3$, ZnO, all that within $550-850$~nm and $390-500$~nm wavelength ranges for Au and Ag, respectively.
The searched intervals for $r_c$ and $t_s$ are justified by the final results shown in Figures~\ref{fig:optAu} and~\ref{fig:optAg} and make use of the fact that too small metal cores cause an increased absorption resulting in relatively enhanced contribution of the nonradiative decay rate which reduces fluorescence.
Dyes are assumed to have zero or a moderate Stokes shift of 25~nm, which are the typical values of Alexa fluorophores. 
%%%%%%%%%%%%%%%%%%%%%%%%%%
%%%%%%%%%%%%%%%%%%%%%%%%%%

In general, larger fluorescence enhancement factor can be achieved with utilizing shells having higher refractive index. 
Interestingly, the maximum fluorescence enhancement of Au and Ag core-shell particles covers a large portion of the visible spectrum. 
For Au cores, the maximum fluorescence enhancement can be achieved between $620$--$700$~nm irrespective of the host (air or water), with the fluorescence enhancement overcoming that around homogeneous metal particle within the first near-infrared biological window (NIR-I) between 700--900~nm by a factor of $\approx 2$, when the shell refractive index $n_s\gtrsim 2$.

Silver has been known for long time as the best surface-enhanced Raman spectroscopy (SERS)~\cite{Moskovits1985} and plasmonic material~\cite{Moroz1999a,Moroz2000}.
An evidence of this is seen also here in almost an order of magnitude higher maximum fluorescence enhancement ($\approx 70$) for Ag cores (Figure~\ref{fig:optAg}) than in the case of Au cores ($\approx 9$) (Figure~\ref{fig:optAu}), which is a consequence of weaker silver losses. This enables stronger enhancement of the near-field electric field intensities and the resulting excitation rate.
Once a proper fluorescence enhancement peak builds up, the ratio of peak fluorescence enhancement values with and without shell is not as high as that for Au cores. However, the difference in the peak values is significantly higher ($\approx 20$) than for Au cores ($\approx 6$). The effect of the Stokes shift on the maximum fluorescence enhancement is clearly visible in both core cases. Given that arbitrary Stokes shifts are, in principle, possible~\cite{Ren2018}, the Stokes shifts could become another useful model parameter.

We conclude this section by an observation, which will be detailed below, that our average fluorescence results shown in  Figures~\ref{fig:optAu} and~\ref{fig:optAg} can be easily exceeded by judiciously placing a low-$q_0$ fluorophore ($q_0\approx 0.1$) at a hot-spot of the core-shell particle, with the fluorophore dipole moment properly oriented, which can generate a multiplication factor of $\gtrsim 25$ on top of the results shown in Figures~\ref{fig:optAu}, \ref{fig:optAg}. 

%\textcolor{blue}{Embedding the core-shell nanoparticles in the aqueous host results in relative small fluorescence enhancements compared to those in the air medium, due to the reduced permitivity contrast between the shell and the host. {\bf Does not apply to Ag! One has to differentiate here. This is why I would put this into the caption of Ag optimization}}
%\textcolor{blue}{Higher fluorescence enhancement factor can be achieved for dyes with a moderate Stokes shift $\approx$ 25~nm, indicating a better balance between the excitation rate and quantum yield. {\bf Again only very minor effect for Au!}}
%\textcolor{blue}{These phenomena are much more pronounced for Ag core, which is attributed to different dielectric function compared to Au.}

%%%%%%%%%%%%%%%%%%%%%%%%%%
\section{Discussion}
\lb{sc:disc}
%%%%%%%%%%%%%%%%%%%%%%%%%%
An essential prerequisite for our simulations was recently reported efficient determination of orientationally averaged electric field intensities~\cite{Rasskazov19JOSAA,stratify}, which supplemented earlier efficient calculation of decay rates~\cite{Moroz2005}. In essence, surface integrals of electric field intensity can be performed analytically~\cite{Rasskazov19JOSAA} and the calculation of average intensity costs the same computational time as determining intensity at a given point.
Below we discuss a number of different aspects related to our optimization results.

\subsection{A critical role of the shell}
\begin{figure}[t!]
    \centering
    \includegraphics{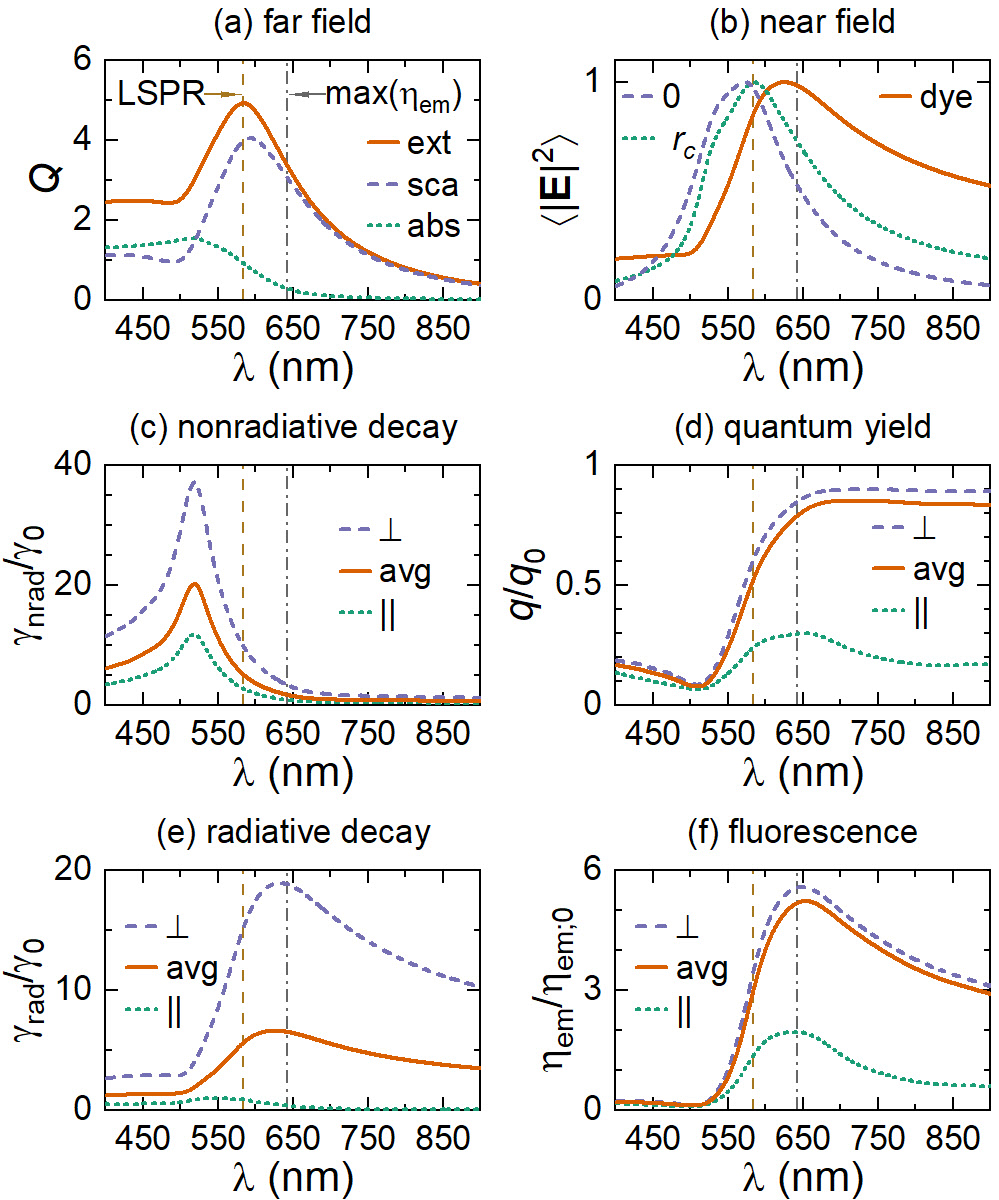}
    \caption{An explanation of the red shift of the maximum fluorescence of Au@SiO${}_2$ core-shell particle with $r_c=70$~nm and $t_s=19.7$~nm tuned to exhibit optimal fluorescence enhancement at $\ld=642$~nm (vertical dash-dot line) in air host relative to its LSPR at $\ld=583$~nm (vertical dashed line):
    (a)~extinction, scattering and absorption efficiencies;
    (b)~orientationally averaged electric field intensity at the center of the Au core (0), at the metal side surface of Au core at $r_c$, and at the dye location at $0.75$~nm distance from the shell surface (dye), each curve is normalized to show the red-shift of the peak moving from the core center to the dye location;
    (c),(e)~nonradiative and radiative decay rates for radial ($\perp$) and tangential ($\parallel$) orientation of the dipole emitter, and orientationally averaged (avg);
    (d)~quantum yield, eq~\ref{eq:QY}, calculated for the respective orientations of the dipole emitter (radial, tangential, and orientationally averaged);
    (f)~fluorescence enhancement, eq~\ref{eq:bff}.
    Except for quantum yield, all other curves exhibit an asymmetric Fano-like shape.
    }
    \label{fig:Au70}
\end{figure}

%%%%%%%%%%%%%%%%%%%%%%%%%%%%%%%
The lesson to be learned from our simulations shown in Figures~\ref{fig:optAu}, \ref{fig:optAg} is obviously that the maximal fluorescence enhancement increases with the shell refractive index. 
This suggests even higher fluorescence enhancements by using shell materials such as Ta${}_2$O${}_5$ ($n_s\approx 2.16$), Nb${}_2$O${}_5$ ($n_s\approx 2.48$), TiO${}_2$ ($n_s\gtrsim 2.55$), or Si ($n_s\gtrsim  3.7$).
To get more insights about the position of, and the mechanism behind, the maximal fluorescence enhancement, we plot in Figure~\ref{fig:Au70} in details near-field (NF) and far-field properties of Au@SiO${}_2$ core-shell with optimal parameters to exhibit maximum achievable fluorescence at $\ld=642$~nm.
An asymmetric Fano-like shape characterizing extinction, scattering and absorption spectra, local electric field enhancement, nonradiative and radiative decay rates, and fluorescence enhancement (Figures~\ref{fig:Au70}a-c,e,f) indicates an interference of different multipolar modes~\cite{Tribelsky2008,Miroshnichenko2010a}. 
As observed by Miroshnichenko~\cite{Miroshnichenko2010a}, the resonant NF enhancement (i.e. local field enhancement that controls the excitation rate, $\gamma_{\rm exc}$) can be {\em red-shifted} in the Fano case by as much as $\approx 50$~nm relative to the resonant scattering (i.e. LSPR) for metal-dielectric core-shell particles. 
The NF red-shift has been known for a long time~\cite{Messinger1981} and its origin can be accounted for by a simple harmonic oscillator model of plasmon oscillations~\cite{Zuloaga2011}.
The maximum of radiative decay rate shown in Figure~\ref{fig:Au70}e coincides with the maximum of NF at the outer shell surface (Figure~\ref{fig:Au70}b), which is seen as the main origin of the {\em red-shift} of the maximum of fluorescence enhancement (Figure~\ref{fig:Au70}f) relative to the LSPR (Figure~\ref{fig:Au70}a). 
An additional small red-shift of the maximum of fluorescence enhancement (Figure~\ref{fig:Au70}f) relative to the maximum of radiative decay rate $\gamma_{\rm rad}$ (Figure~\ref{fig:Au70}e) is due to increasing quantum yield (Figure~\ref{fig:Au70}d).
Note in passing that it has been noted for homogeneous Ag particles that the highest fluorescence enhancement is obtained for an emission wavelength red-shifted from the LSPR~\cite{Bharadwaj2007}, yet no explanation has been given.

Of the same magnitude as the red-shift is the {\em blue shift} of the maximum of nonradiative decay rate (Figure~\ref{fig:Au70}c) relative to the LSPR (Figure~\ref{fig:Au70}a). 
The LSPR position coincides with the maximum of field intensity at the Au core surface (Figure~\ref{fig:Au70}b) and the maximum of $Q_{\rm ext}$ (Figure~\ref{fig:Au70}a). 
The position of the maximum nonradiative decay rate $\gamma_{\rm nrad}$, which essentially coincides with $Q_{\rm abs}$, is blue-shifted to the LSPR.
The difference in the shape of the peaks of $\gamma_{\rm nrad}$ and $Q_{\rm abs}$ is attributed to a fact that the former describes power loss of a dipole emitter at the shell surface, whereas the latter is the measure of absorption of incident plane arriving from the spatial infinity.
The above red and blue shifts relative to the LSPR of corresponding optimized core-shell particle provide important design rules for selected application. 
Note is passing that there is a number of applications where not only fluorescence \textit{enhancement} but also an efficient fluorescence \textit{quenching} is highly desirable~\cite{Dubertret2001}.

%%%%%%%%%%%%%%%%%%%%%%%%%%%%%%%%%%%%%%%%%%
\begin{figure}
    \centering
    \includegraphics{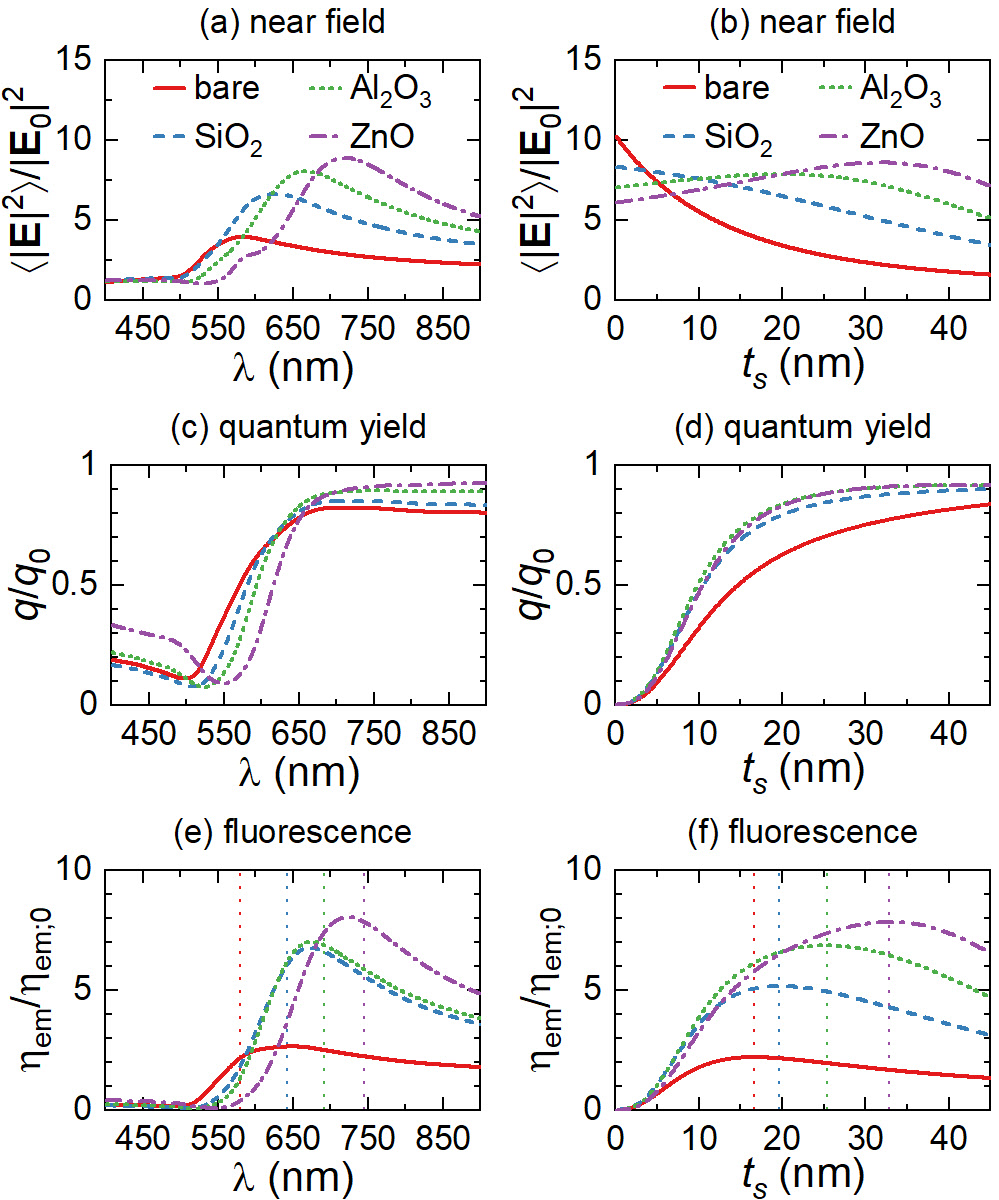}
    \caption{The impact of shells from different dielectrics on the fluorescence enhancement of $r_c=70$~nm Au nanoparticle in air host.
    Wavelength- (left) and shell-thickness- (right) dependent: 
    (a),(b)~orientationally averaged electric field intensity at the dye location at $0.75$~nm distance from the shell surface;
    (c),(d)~quantum yield, eq~\ref{eq:QY};
    (e),(f)~fluorescence enhancement, eq~\ref{eq:bff}.
    Vertical dotted lines in (e),(f) show the respective wavelengths and corresponding optimal shell thicknesses to reach maximum fluorescence enhancement for bare Au core ($\ld=580$~nm and $t_s=16.7$~nm), Au@SiO${}_2$ ($\ld=642$~nm and $t_s=19.7$~nm), Au@Al${}_2$O${}_3$ ($\ld=691$~nm and $t_s=25.4$~nm), Au@ZnO ($\ld=745$~nm and $t_s=32.8$~nm).
    Wavelength-dependent plots (left) are presented for optimal shell thicknesses denoted by vertical lines in (f), while shell-thickness-dependent plots (right) show corresponding quantities at wavelengths denoted by vertical lines in (e).
    }
    \label{fig:Au70_shells}
\end{figure}

%%%%%%%%%%%%%%%%%%%%%%%%%%%%%%%%%%%%%%%%%%
In a Drude-like region above bulk plasma wavelength ($\ld_p\approx 328$~nm for Ag and $\ld_p\approx 226$~nm for Au), the loss tangent (i.e. the ratio $|\mb{Im }\veps/\mb{Re }\veps|$) of a noble metal dielectric function decreases with increasing wavelength.
Therefore, the use of shells with larger refractive indices, which increases the red shift of their LSPR, and consequently of their NF maximum, creates increasingly favourable conditions for fluorescence enhancement by reducing (increasing) the overall contribution of nonradiative (radiative) decay rates. 
At the same time, the boundary condition at the shell-host interface for the radial components of electric field, $\veps_s \vE_s^\perp= \veps_h  \vE_h^\perp$, implies that $\vE^\perp$ experiences a jump by the factor of $\veps_s/\veps_h$ at the host, with $\vE^\parallel$ being continuous across the shell-host interface. 
Obviously, the jump in field intensity values at the shell surface becomes more pronounced with larger $n_s$, enabling to achieve quite large electric field enhancement at the shell-host interface even for relatively thick shells (Figure~\ref{fig:Au70_shells}b; Figures S3, S4, see Supporting Information).

Note that the procedure used in Figures~\ref{fig:optAu} and \ref{fig:optAg} for the search of optimal ($r_{c;{\rm opt}},t_{s;{\rm opt}}$) configurations implies the fluorescence enhancement to be the largest possible at a given wavelength $\ld_{\rm exc;opt}$.
This, however, does not restrict the same core-shell to exhibit even larger fluorescence enhancement at another wavelength $\ld_{\rm exc}\neq \ld_{\rm exc;opt}$, which is clearly observed in Figure~\ref{fig:Au70_shells}e.
Nonetheless, the maxima in Figure~\ref{fig:Au70_shells}f, which perfectly correspond to $t_{s;{\rm opt}}$, confirm that ($r_{c;{\rm opt}},t_{s;{\rm opt}}$) configurations indeed provide optimal $\eta_{\rm em}/\eta_{\rm em;0}$ at $\ld_{\rm exc;opt}$ (cf. Figure S2, see Supporting Information).
%%%%%%%%%%%%%%%%%%%%%%%%%%%%%%%%%%%%%%%%%%
%%%%%%%%%%%%%%%%%%%%%%%%%%%%%%%%%%%%%%%%%%

\subsection{Comparison with an experiment}
\lb{sc:exp}
%%%%%%%%%%%%%%%%%%%%%%%%%%%%%%%%%%%%%%%%%%
Our optimization results of Sec.~\ref{sc:optim} provide general guidelines for achieving the highest possible fluorescence enhancements.
Before comparing them against an experiment, the following has to be taken into account. 
First, our simulations assuming either zero Stokes shift, or a model Stokes shift of $25$ nm have to be adjusted to the Stokes shift of the fluorophore used. 
The Stokes shift is strongly dependent on a fluorophore used and can be much larger than the $25$ nm~\cite{Ren2018}. 
Second, optimally fabricated Au and Ag cores may exhibit lower losses~\cite{McPeak2015} than those following from the data of Palik et al~\cite{Palik1998} used in our simulations, thereby facilitating even higher fluorescence enhancements. 
Third, optimization results of Sec.~\ref{sc:optim} assumed orientationally averaged electric field intensity and average dipole orientation. 
On judiciously placing fluorophore at hot spots of electric field intensity with the fluorophore dipole moment properly oriented such a local fluorescence enhancement can be up to $\approx 2.5$ times stronger (Figure S9, see Supporting Information).
Last but not the least, simulations of Sec.~\ref{sc:optim} presumed intrinsic quantum yield $q_0=1$. 
In this regard, any $q_0<1$ will increase the denominator of eq~\ref{eq:QY} by $(1-q_0)/q_0$. 
However, for moderate $0.1 \le q_0<1$ and for the metal-dye separations studied, this contribution is typically negligible compared to other two terms in the denominator (e.g. $\gamma_{\rm nrad}/\gamma_{{\rm rad};0}$). 
Hence $0.1 \le q_0<1$ will hardly change the resulting quantum yield $q$ of eq~\ref{eq:QY}.
However, such a moderate $0.1 \le q_0<1$ will, according to eq~\ref{eq:bff}, significantly increase the resulting fluorescence enhancement shown in Figures~\ref{fig:optAu}, \ref{fig:optAg} by a factor of $1/q_0$. 
For instance, in the case of low $q_0=0.36$ carboxyfluorescein (FAM) and Au@SiO${}_2$ core-shell nanoparticles with $r_c=29.5$ nm of Ref.~\citenum{Tovmachenko2006} [Table 1; Figure 2B] this amounts to the factor of $1/q_0\approx 2.8$.
On the other hand, for cascade yellow (CYe) having an intrinsic quantum efficiency of about $q_0=0.56$ and Ag@SiO${}_2$ core-shell nanoparticles of Ref.~\citenum{Tovmachenko2006} [Table 1; Figure 3] this amounts to the factor of $1/q_0\approx 1.78$. 
Therefore, when comparing with experiment, a low-$q_0$ dye ($q_0\approx 0.1$) placed at a hot-spot of the core-shell particle can easily generate a multiplication factor of $\approx 25$ by which the results shown in Figures~\ref{fig:optAu}, \ref{fig:optAg} are to be multiplied. 
Not surprisingly, initial experiments observed significant fluorescence enhancements especially with low-$q_0$ dyes~\cite{Tovmachenko2006,Aslan2007}.

\subsection{Comparison with metal shells}
\lb{sc:wmsh}
%%%%%%%%%%%%%%%%%%%%%%%%%%%%%%%%%%%%%%%%%%
The highest reported fluorescence enhancements for dielectric ($n=3.5$) core and Ag shell of comparable sizes tuned to the Fano resonance~\cite{Arruda2017a} were in the case of radially oriented dipole located at the hot spot reaching nearly the value of $50$, whereas the enhancements for tangentially oriented dipole hardly exceeded the value of $2.5$. This results in the dipole orientation averaged fluorescence enhancement of $\approx 18$.
Our results for dielectric shells presented in Figure~\ref{fig:optAg}
show almost {\em four-times} larger averaged fluorescence enhancement with much lower shell refractive index ($n=2$) and without the need of keeping the dipole emitter at a hot spot. Here one notable difference between the homogeneous sphere and metal-dielectric core-shell on one hand, and a dielectric-metal core-shell on the other hand, is that in the former case the areas of highest field enhancement are located near the particle poles on the rotation axis {\em parallel} to the incident polarization direction, whereas in metallic shells the areas of highest field enhancement are located near the particle poles on the rotation axis {\em perpendicular} to the incident polarization~\cite{Schelm2005a}.

\subsection{Quasi-static approximation}
\begin{figure}
    \centering
    \includegraphics{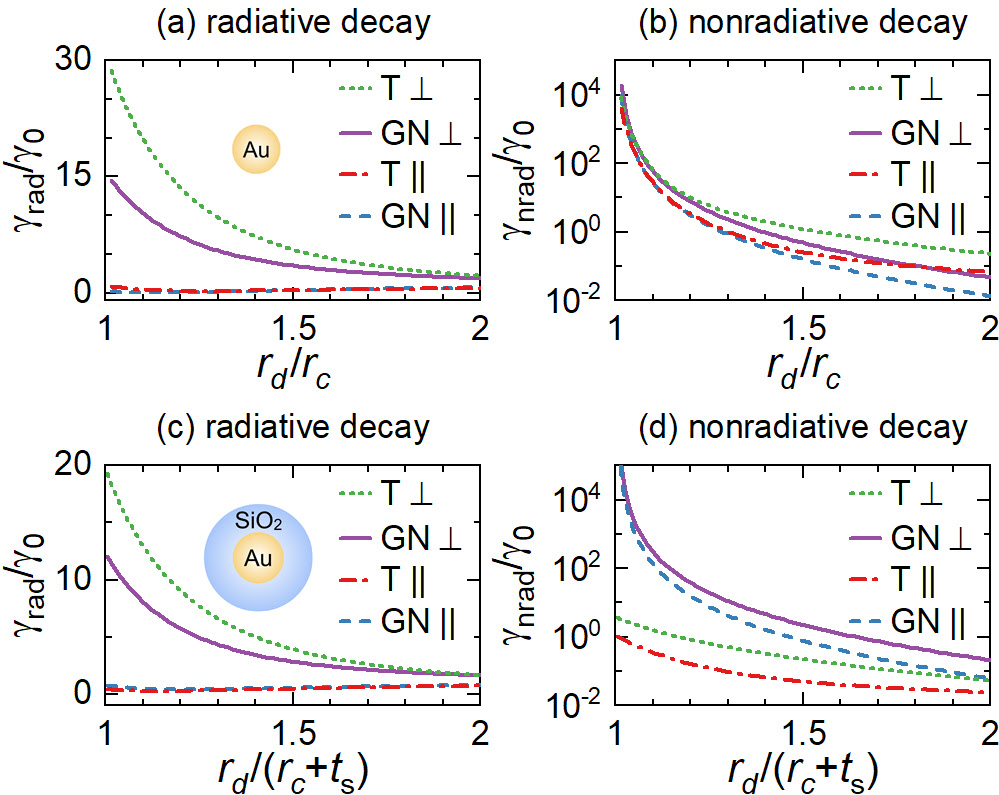}
    \caption{Radiative (left) and nonradiative (right) decay rates at a distance $r$ from a center of nanoparticle, calculated via transfer-matrix method (T) and the Gersten and Nitzan (GN) quasi-static approximation for bare Au nanoparticle (top) and for Au@SiO${}_2$ core-shell for radial ($\perp$) and tangential $(||)$ orientation of the dipole emitter.
    The parameters are given in Figure~\ref{fig:Au70_shells} caption.
    }
    \label{fig:Au70_GN}
\end{figure}

\lb{sc:GN}
%%%%%%%%%%%%%%%%%%%%%%%%%%%%%%%%%%
The Gersten and Nitzan (GN) quasi-static approximation for determining decay rates \cite{Gersten1981}, which makes use of particle multipolar polarizabilities, $\al_\ell$, has been known to provide a very good approximation of radiative and nonradiative decay rates in the case of small homogeneous particles \cite{Moroz2010}.
A dipolar polarizability with the account of dynamic depolarization and radiative correction \cite{Moroz2009} (see Sec. 3 in Supporting Information for details) enables rather precise description of the scattering properties of core-shell particles. 
Nevertheless, the GN approximation~\cite{Gersten1981} largely fails to describe the nonradiative decays rates for the core-shell configurations studied here, as shown in Figure~\ref{fig:Au70_GN}d. 
Although the modified long wavelength approximation (MLWA) and its variants \cite{Moroz2009,Chung2009} capture well the far-field properties (e.g. scattering), they do not perform so well in capturing the near-field properties (e.g. fluorescence). 
An indirect indication of this is that the radiative decay rates, requiring only dipole polarizability in the GN approximation, were approximated much better than the nonradiative decay rates, which require all multipole polarizabilities. 

%%%%%%%%%%%%%%%%%%%%%%%%%%
%%%%%%%%%%%%%%%%%%%%%%%%%%

%%%%%%%%%%%%%%%%%%%%%%%%%%
\section{Conclusion}
\lb{sc:concl}
%%%%%%%%%%%%%%%%%%%%%%%%%%
Large scale simulations were performed by means of the transfer-matrix method to reveal optimal conditions for metal-dielectric core-shell particles to induce an optimal fluorescence of a fluorophore on their surface. 
In the simulations, limited to Au and Ag cores and common dielectric shell materials (SiO${}_2$, Al${}_2$O${}_3$, ZnO), we have (i) determined the optimal size of metal core and shell thickness for reaching a maximum fluorescence enhancement for each emission wavelength, and then (ii) determined overall maximum fluorescence enhancements taken over entire wavelength interval. The peak value of maximum achievable fluorescence enhancement factors of core-shell nanoparticles can reach up to 9 or 70 for Au and Ag cores within $600-700$~nm and $400-450$~nm wavelength ranges, respectively, which is much larger than that for corresponding homogeneous metal nanoparticles. 
Replacing air by an aqueous host has a dramatic effect of nearly halving the sizes of optimal core-shell configurations at the maximum of achievable fluorescence. 
In the case of Au cores, the fluorescence enhancements for wavelengths within the first near-infrared biological window (NIR-I) between 700 and 900~nm can be improved twofold compared to homogeneous Au particle when the shell refractive index $n_s\gtrsim 2$. Given that the maximum achievable fluorescence enhancement factor increases with the shell refractive index, even higher fluorescence enhancements could be possible by using shell materials such as Ta${}_2$O${}_5$ ($n_s\approx 2.16$), Nb${}_2$O${}_5$ ($n_s\approx 2.48$), TiO${}_2$ ($n_s\gtrsim 2.55$), or Si ($n_s\gtrsim  3.7$).
As a rule of thumb, the wavelength region of optimal fluorescence (maximal nonradiative decay) turned out to be red-shifted (blue-shifted) by as much as $50$~nm relative to the LSPR of corresponding optimized core-shell particle.
The main contribution to the red-shift was determined to be provided by the red shift of the near-field enhancement relative to the LSPR. 
Our results provide important design rules and general guidelines for enabling versatile platforms for selected applications such as imaging, light source, and biological applications. Our results for near-field enhancement also have direct relevance to designing optimal SERS platforms.

%%%%%%%%%%%%%%%%%%%%%%%%%%
\begin{acknowledgement}
CAEP Innovation Grant, PRC China (Grant No.CX20200011), Science Challenge Project, PRC China (Grant No.TZ2016003).
\end{acknowledgement}
%%%%%%%%%%%%%%%%%%%%%%%%%%

%%%%%%%%%%%%%%%%%%%%%%%%%%
\begin{suppinfo}
Benchmarking with the finite-element method; additional data for extinction spectra, electric field enhancement, radiative and nonradiative decay rates, quantum yields and fluorescence enhancement for core-shell nanoparticles; quasi-static equations.
\end{suppinfo}
%%%%%%%%%%%%%%%%%%%%%%%%%%

%%%%%%%%%%%%%%%%%%%%%%%%%%
%\section*{Disclosures}
%%%%%%%%%%%%%%%%%%%%%%%%%%
%The authors declare no conflicts of interest.

%%%%%%%%%%%%%%%%%%%%%%%%%%
\bibliography{references}
%%%%%%%%%%%%%%%%%%%%%%%%%%

\end{document}

\section{Additional data on calculation methods}

\subsection{Benchmarking with the finite-element method}
\begin{figure}[t]
	\centering
	\includegraphics{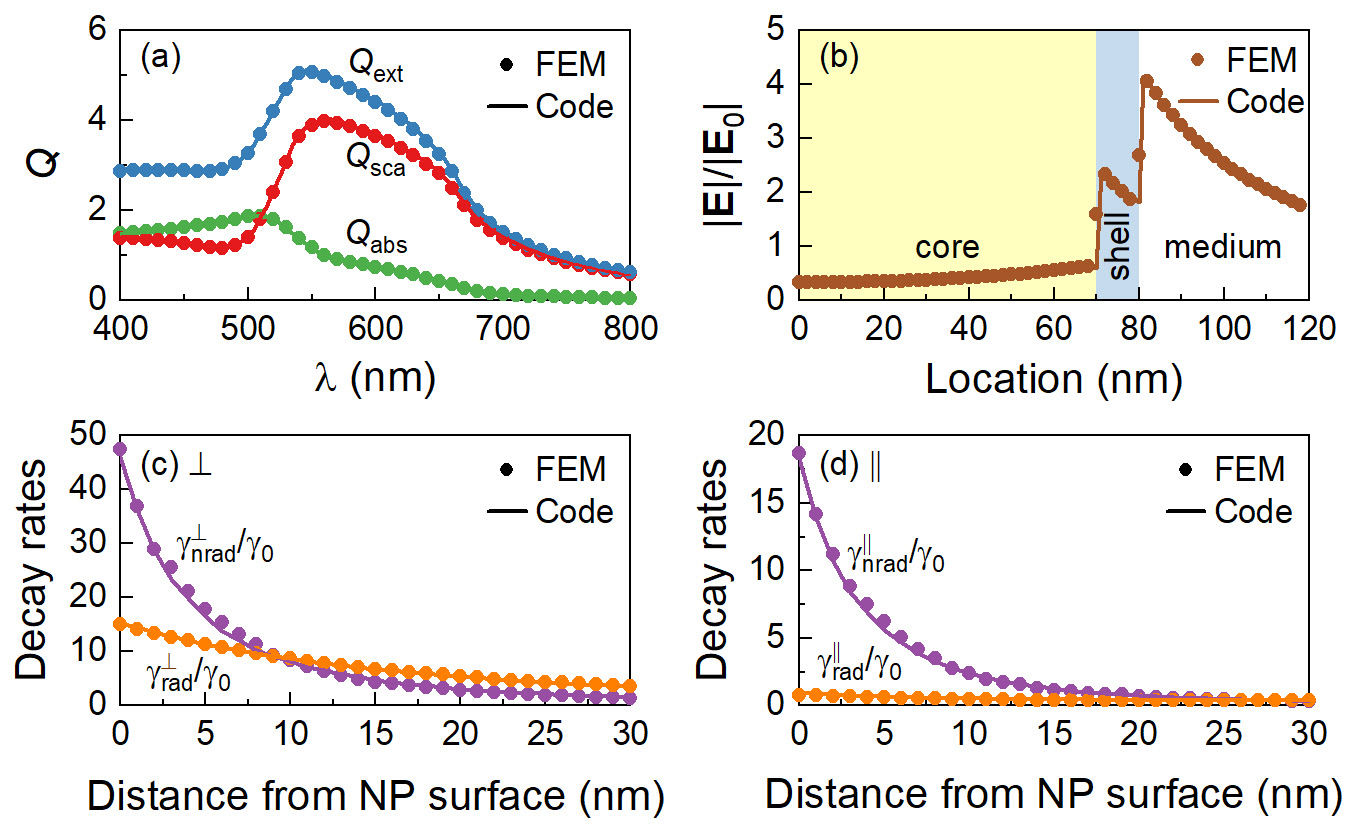}
    \caption{
    Benchmark results for Au@SiO${}_2$ core-shell:
    (a) extinction, scattering and absorption spectra, 
    (b) normalized electric field distributions under plane wave excitation, and fluorescence decay rates for 
    (c) perpendicular and 
    (d) parallel dipole orientation. 
    Parameters are set as: core radius $r_c=70$~nm, shell thickness $t_s=10$~nm, shell refractive index $n_s=1.45$ at wavelength $\lambda=550$~nm.
    A single FEM simulation takes 2-3 hours for fine mesh on the HPC server with 128 CPU and 512 GB RAM used. In contrast, 
    our transfer matrix code based on analytical solution of Maxwell equations runs few seconds on a regular workstation with 16 CPU and 64 GB RAM.}
    \label{fig:bnchmrk}
\end{figure}

\lb{sc:bnch}
%%%%%%%%%%%%%%%%%%%%%%%%%%
Our transfer matrix method~\ct{Moroz2005,Rasskazov19JOSAA} is an exact analytic solution of the Maxwell equations.
The analytical methods from Sec. 2 have been implemented with the in-house developed Matlab code~\cite{stratify}, which has been used in simulations. 
Its Fortran version for decay rates has been since 2006 freely available from 
\verb|www.wave-scattering.com/codes.html|.
In order to disperse any doubt on the validity of our method we have found it expedient to compare it against the finite element method (FEM) using COMSOL Multiphysics, for the case of Au@SiO${}_2$ core-shell.
In FEM simulations, the background field was set to be the plane wave during the excitation process and zero during the dipole emission process. 
The perfect matching layer (PML) surrounding the computational domain was used to absorb any unphysical reflections. 
The near field distribution can be probed anywhere in the computational domain, and the normalized fluorescence decay rates can be obtained straightforwardly using the scattering and absorption power quantities~\cite{Ma2018,Sun2018,Sun2019a}. 
The benchmark results are illustrated in Figure~\ref{fig:bnchmrk}. 
It is clear that our code agrees with the FEM simulations. 
However our transfer matrix code, which was specifically designed for core-shell particles, is several orders of magnitude faster.

%%%%%%%%%%%%%%
%%%%%%%%%%%%%%

\subsection{Finding optimal core-shell configurations}
\lb{sc:opt}
%%%%%%%%%%%%%%%%%%%%%%%%%%

\begin{figure}
    \centering
    \includegraphics{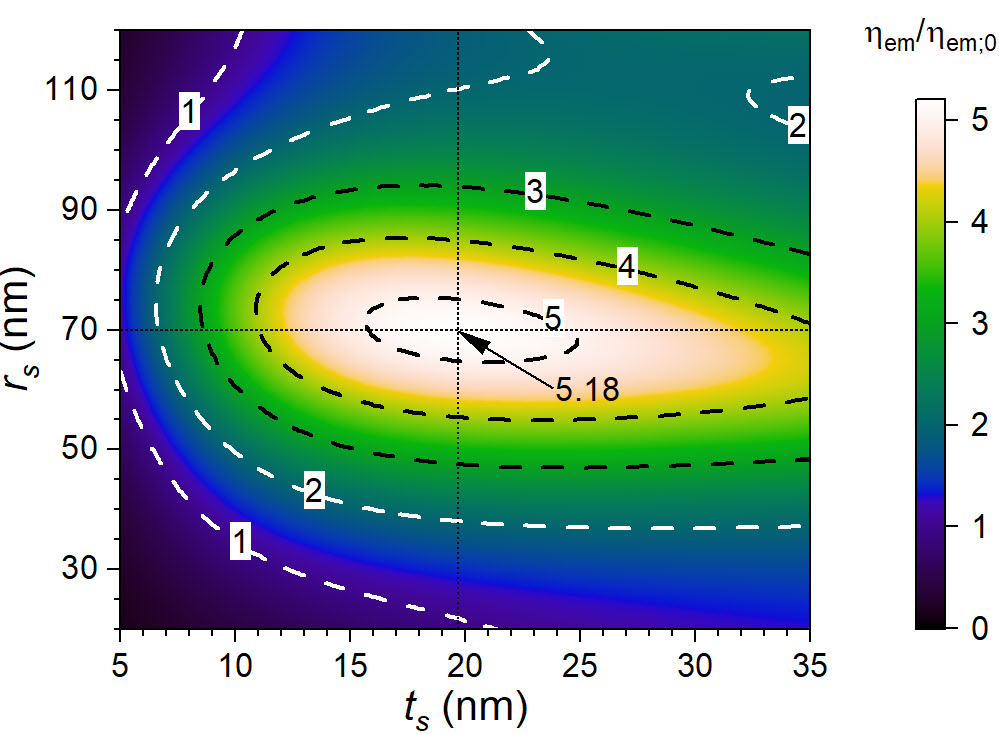}
    \caption{Fluorescence enhancement at $\ld_{\rm exc} = \ld_{\rm em} = 642$~nm for Au@SiO${}_2$ core-shell nanoparticle embedded in air. 
    The ${\rm max}(\eta_{\rm em}/\eta_{\rm em;0})=5.18$ is observed for $r_c=70$~nm and $t_s=19.7$~nm which are shown with black horizontal and vertical dotted lines, correspondingly. 
    }
    \label{fig:opt}
\end{figure}

Figure~\ref{fig:opt} shows the example of $\eta_{\rm em}/\eta_{\rm em;0}$ calculation in $(r_c,t_s)$ parameter space for Au@SiO${}_2$ core-shells at $\ld_{\rm exc} = \ld_{\rm em} = 642$~nm embedded in air (cf. Figures 4 and 5).
A clear existence of an optimal core-shell configuration for gaining maximum fluorescence enhancement is observed.
Noteworthy, the moderate deviations of $r_c$ and $t_s$ from optimally chosen values may yield in only slight decrease of the fluorescence enhancement factor.

%%%%%%%%%%%%%%%%%%%%%%%%%%
\section{Additional data for fluorescence enhancement}
%\section{Fluorescence enhancement by Au@dielectric core-shell particle}
\lb{sc:Au_shell}
%%%%%%%%%%%%%%%%%%%%%%%%%%

\subsection{Electric field enhancement at shell-medium interface}
\begin{figure}[t]
	\centering
	\includegraphics{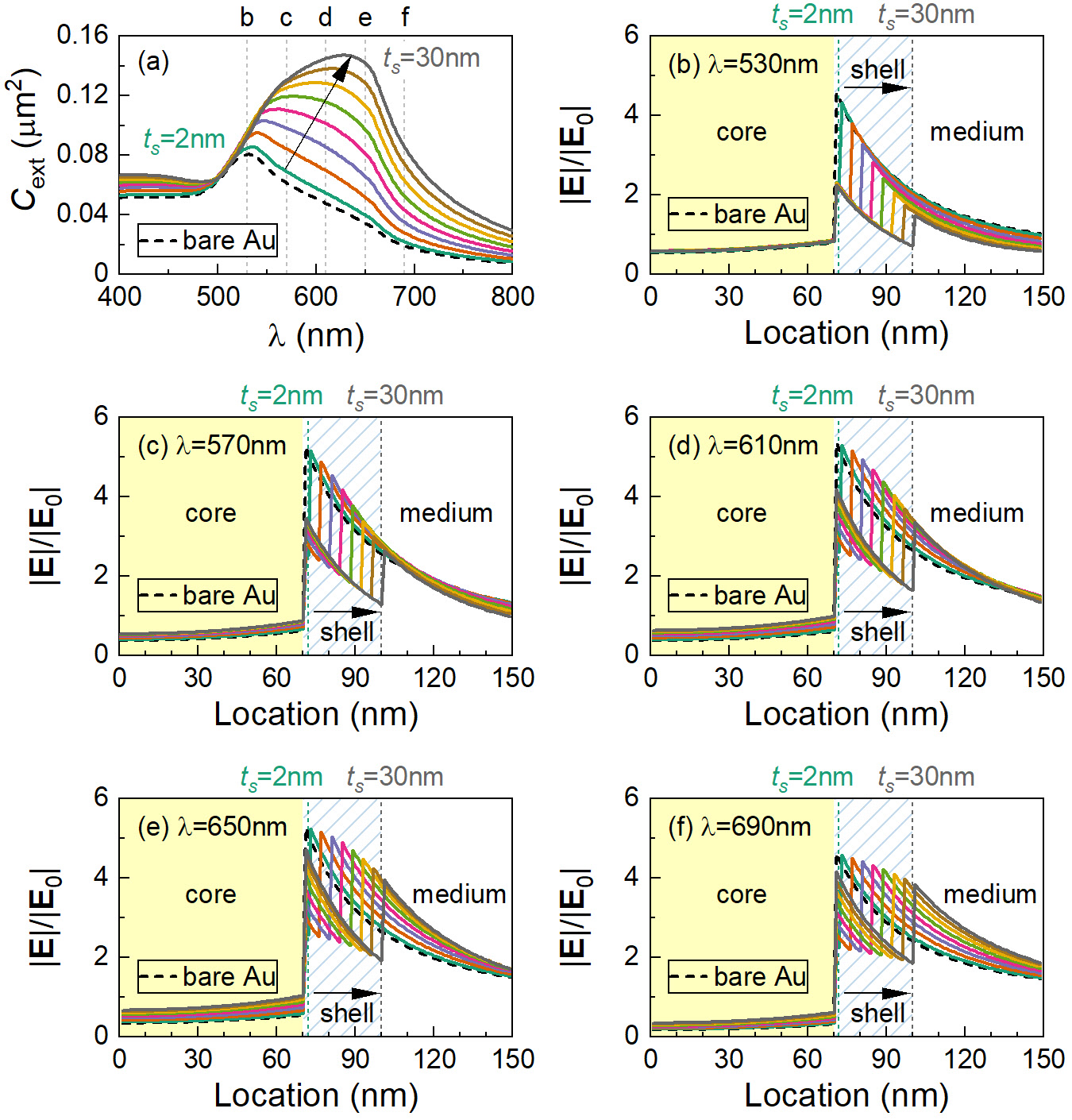}
    \caption{
    (a) Extinction spectra of core-shell nanoparticle with shell thicknesses from $t_s=2$~nm to $t_s=30$~nm with a 4~nm step. 
    (b)-(f) Electric field enhancement at different wavelengths. 
    Parameters are set as: core radius $r_c=70$~nm, shell refractive index $n_s=1.45$.
    Black dashed lines show corresponding curves for bare Au nanoparticles.}
    \label{fig:extflds}
\end{figure}

%%%%%%%%%%%%%%%%%%%%%%%%%%
To elaborate the optimized results in Figures 2 and 3, below we discuss the effect of dielectric shell in manipulating various fluorescence parameters. 
For illustration purpose, Au@dielectric structure is considered, while the same conclusion can be extended to the Ag core cases. 
The emitter is assumed to be located at the hot spot of the shell surface for both perpendicular and parallel emitter dipole orientations.
Under the plane wave excitation, it is well known that the presence of the dielectric shell red-shifts the LSPR~\cite{Neeves1989} as shown in Figure~\ref{fig:extflds}a, where the shell thickness $t_s$ varies from 0~nm to 30~nm, and the shell refractive index $n_s=1.45$ (SiO${}_2$).
The corresponding resonant peak position changes from 530~nm (homogeneous Au sphere) to 630~nm ($t_s= 30$~nm). 
The evolution of electric field distributions of the core-shell nanoparticle at different wavelengths ($\lambda = 530, 570, \dots , 690$~nm) are illustrated in Figure~\ref{fig:extflds}b--\ref{fig:extflds}f, respectively. 
The electric field intensity increases at the core/shell interface with subsequent decrease inside the dielectric shell with increasing distance to the core. 

%The boundary condition at the shell-host interface for the radial components of electric field, $\veps_s \vE_s^\perp= \veps_h  \vE_h^\perp$, implies that $\vE^\perp$ experiences a jump by the factor of $\veps_s/\veps_h$ at the host.
%Given that $\vE^\parallel$ is continuous across the shell-host interface, the above jump explains why the electric field intensity in the surrounding host medium is higher than inside the shell for $\veps_s > \veps_h$. 
At the LSPR of a homogeneous Au sphere at $\lambda=530$~nm, the electric field intensity of the core-shell particles are generally below that of a homogeneous Au sphere, regardless of the shell thickness $t_s$, because the LSPR of core-shell particles is red shifted.
With increasing distance from the shell, electric field intensity gradually decreases to the incident field intensity value.
As $\lambda$ increases, the electric field intensity at the shell-medium interface can achieve higher values than that at the same distance from Au surface of a homogeneous particle. 
In addition, the increment of the electric field at the shell-medium interface is more prominent for a larger $t_s$ at a larger $\lambda$, since the electric field strength and the resonant wavelength are inter-correlated.

%%%%%%%%%%%%%%
%%%%%%%%%%%%%%

\begin{figure}[t]
    \centering
    \includegraphics{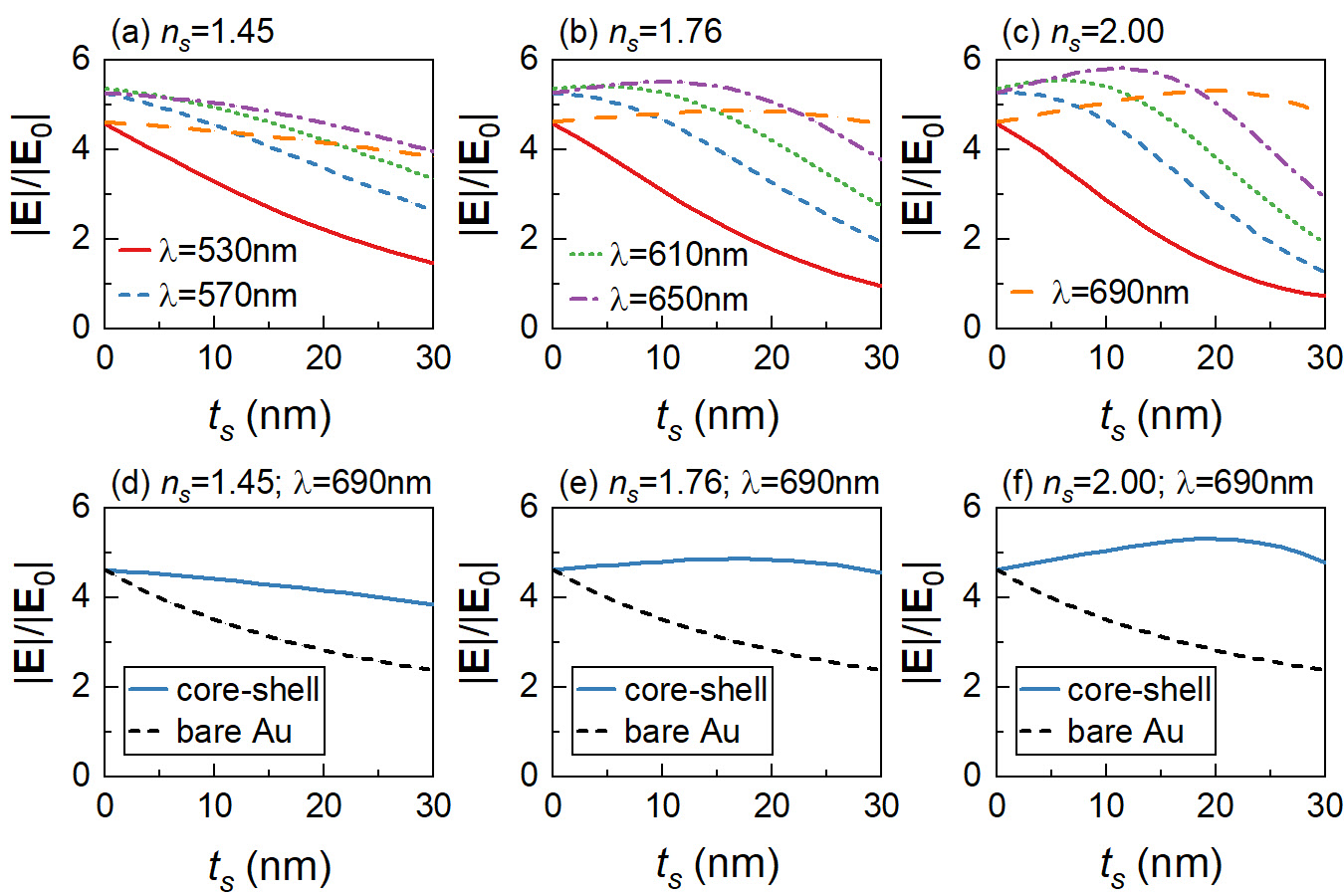}
    \caption{(a)-(c) Electric field  intensity at the shell-medium interface as a function of shell thicknesses for shell refractive indices $n_s=1.45$, 1.76 and 2.00, respectively. 
    (d)-(f) Comparison between electric field at shell-medium interface and that of Au sphere in homogeneous medium at wavelength $\lambda = 690$~nm.
    Au core radius is $r_c=70$ nm in all cases.
    Dielectric shell with higher refractive index could boost electric field at shell-medium interface more effectively, and slow its decay even further.}
    \label{fig:flds}
\end{figure}

Figure~\ref{fig:flds}a shows the corresponding electric field intensity at the shell-medium interface as a function of the shell thicknesses. 
It is found that the dielectric shell can slow the decrease of electric field at the shell-medium interface, in particular for a large wavelength. 
For illustration, Figure~\ref{fig:flds}d compares the profiles of the electric fields at the shell-medium interface with those of Au sphere in the homogeneous medium at $\lambda=690$~nm. 
It is obvious that the electric fields at the shell-medium interface decrease slower than that of the Au sphere in the homogeneous medium. 
The electric field at $t_s=30$~nm drops only by 16.8\% compared to that at $t_s=0$~nm, whereas the electric field around Au sphere drops by 48.5\% in the homogeneous medium at the $30$~nm distance.

%%%%%%%%%%%%%%
%%%%%%%%%%%%%%

In Figure~\ref{fig:flds}a, the highest electric field always occurs at $t_s=0$~nm regardless of the wavelength. 
Such constraint can easily be broken by utilizing the dielectric shell with higher refractive index. 
Figures~\ref{fig:flds}b-c monitor the electric fields at the shell-medium interface with relatively larger shell refractive indices $n_s=1.76$ (Al${}_2$O${}_3$) and $n_s=2.00$ (ZnO), respectively. A high shell refractive index could boost the electric field at the shell-medium interface more effectively, leading to a larger electric field than that at $t_s=0$~nm, which in turn could further slow its decrease. 
Figures~\ref{fig:flds}e-f confirm this point by comparing the profiles of the electric fields at the shell-medium interface with that of Au sphere in a homogeneous medium at $\lambda=690$~nm. 
The electric field decreases even more slower than that with $n_s=1.45$ (see Figure~\ref{fig:flds}e), i.e. the electric field at $t_s=30$~nm drops by merely 1.4\% compared to that at $t_s=0$~nm for $n_s=1.76$, and it even increases by 3.1\% for $n_s=2.00$.
The electric field at the shell-medium interface could be maintained at a high value even with a relatively thick dielectric shell. If the emitter is attached at the core-shell surface, it will experience a much higher fluorescence excitation rate than around a Au sphere at an equivalent distance. 
Combined with a decelerated decrease of electric field, a better balance between the fluorescence excitation rate and quantum yield can be achieved.

\subsection{Radiative and non-radiative decay rates at shell-medium interface}
%%%%%%%%%%%%%%%%%%%%%%%%%%

\begin{figure}
	\centering
		\includegraphics{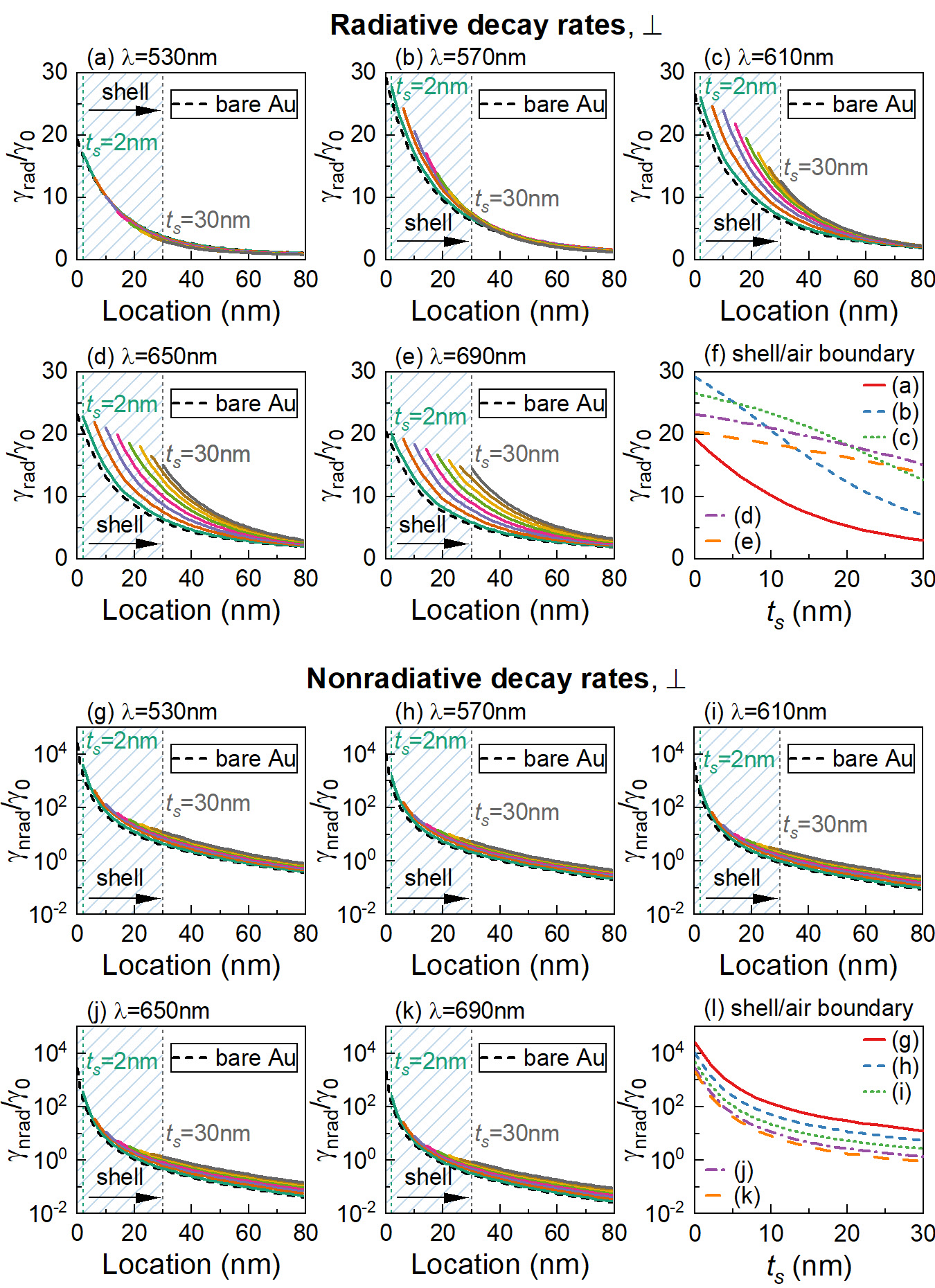}
	\caption{
	(a)-(e) Radiative and (g)-(k) non-radiative decay rates of radially oriented dipole emitter as a function of the distance from nanoparticle surface for various wavelengths. 
	(f) Radiative and (l) non-radiative decay rates at the shell-medium interface, respectively. Au core radius is $r_c=70$ nm in all cases.
	Core-shell structure could manipulate both radiative and non-radiative decay rates, leading to decelerated decrease at shell-medium interface.
	}
	\label{fig:decprpdist}
\end{figure}

The presence of the dielectric shell could not only manipulate the excited electric field distribution, but also has a remarkable influence on the dipole decay rates as governed by eq~2. 
Figures~\ref{fig:decprpdist}a--\ref{fig:decprpdist}e and \ref{fig:decprpdist}g--\ref{fig:decprpdist}k show the radiative and non-radiative decay rates of a perpendicular-oriented emitter. 
The results confirm that the core-shell structure could significantly manipulate both the radiative and non-radiative decay rates at the shell-medium interface, leading to the decelerated decrease as shown in Figure~\ref{fig:decprpdist}f and \ref{fig:decprpdist}l, respectively.
Again, the dipole decay rates at the shell-medium interface decrease slower at a larger wavelength, where the thicker dielectric shell has stronger influence.

%%%%%%%%%%%%%%
%%%%%%%%%%%%%%

\begin{figure}
	\centering
	\includegraphics{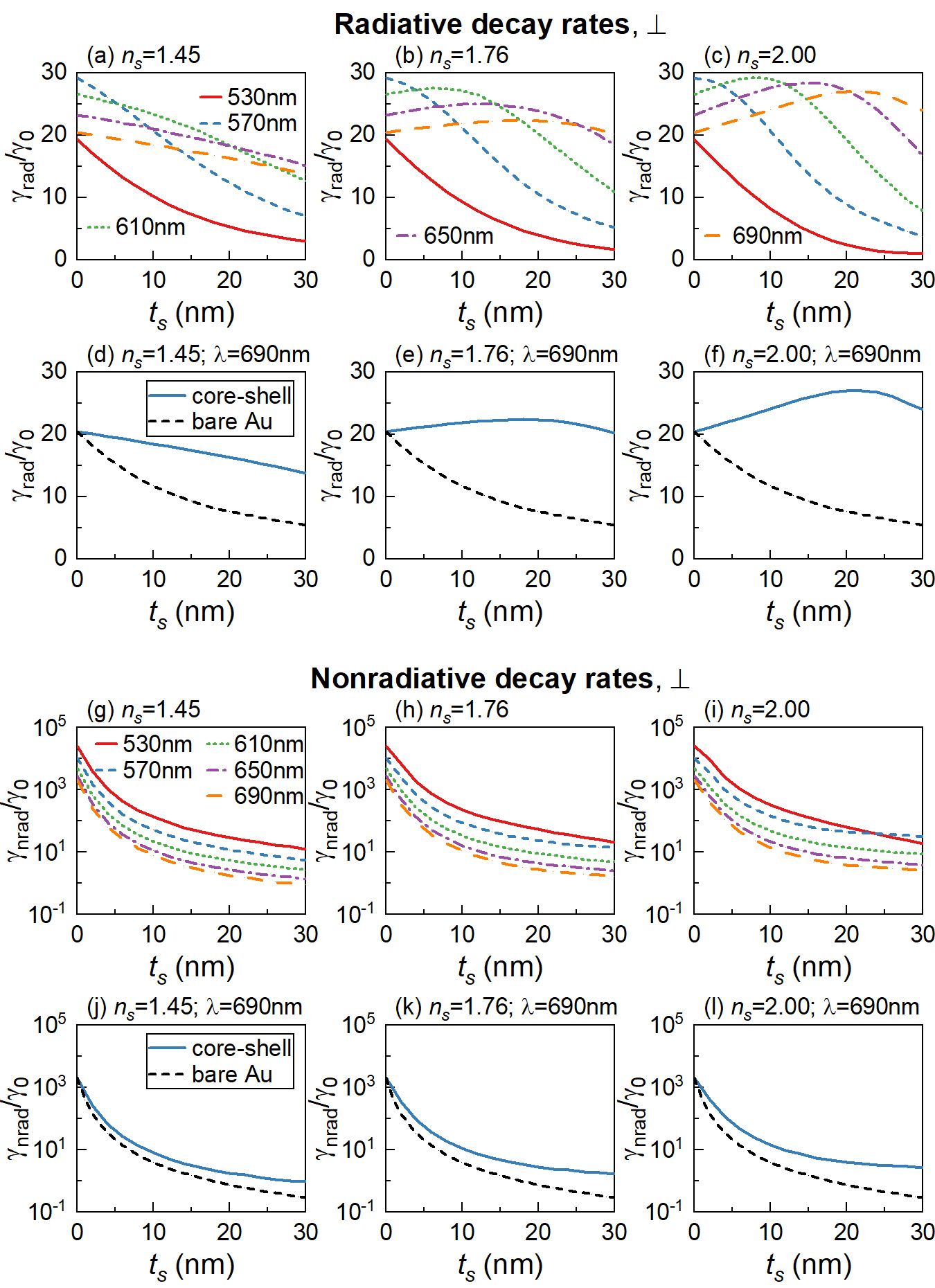}
	\caption{
	(a)-(c) Radiative and (g)-(i) non-radiative decay rates of perpendicular-oriented emitter at the shell-medium interface for shell refractive indices $n_s=1.45$, 1.76 and 2.00. 
	Comparison of (d)-(f) radiative and (j)-(l) non-radiative decay rates at the shell-medium interface with those of Au sphere in homogeneous medium for $\lambda = 690$~nm.
	Au core radius is $r_c=70$ nm in all cases.
	Dielectric shell with higher refractive index could improve both radiative and non-radiative decay rates at the shell-medium interface more effectively, and to slow their decrease even further.}
	\label{fig:decprpshll}
\end{figure}

Likewise, the decelerated decrease of the dipole decay rates at the shell-medium interface can be further manipulated by varying the refractive index of the dielectric shell. 
Figures~\ref{fig:decprpshll}a-c and \ref{fig:decprpshll}g-i show dipole decay rates at the shell-medium interface for a {\em radially} oriented dipole emitter in the case of the shell refractive indices $n_s=1.45$, 1.76 and 2.00, respectively. 
The comparison of these quantities to those of the Au sphere in the homogeneous medium for $\lambda=690$~nm are illustrated in Figure~\ref{fig:decprpshll}d--\ref{fig:decprpshll}f and \ref{fig:decprpshll}j--\ref{fig:decprpshll}l, respectively. 
The results show that the dielectric shell with higher refractive index could further slow the decrease of 
both radiative and non-radiative decay components at the shell-medium interface.
In particular, the highest radiative decay rates at the shell-medium interface with a shell thickness $t_s=20$~nm 
and a shell refractive index $n_s=1.76$ and 2.00 could easily exceed those for Au sphere. 
Note that both radiative and non-radiative decay rates could be enhanced at the shell-medium interface compared 
to those of the homogeneous Au sphere, because both the absorption and scattering efficiency of core-shell nanoparticle are larger than those of the homogeneous Au sphere~\cite{Reineck2013}.

%%%%%%%%%%%%%%
%%%%%%%%%%%%%%

%%%%%%%%%%%%%%
\begin{figure}
	\centering
	\includegraphics{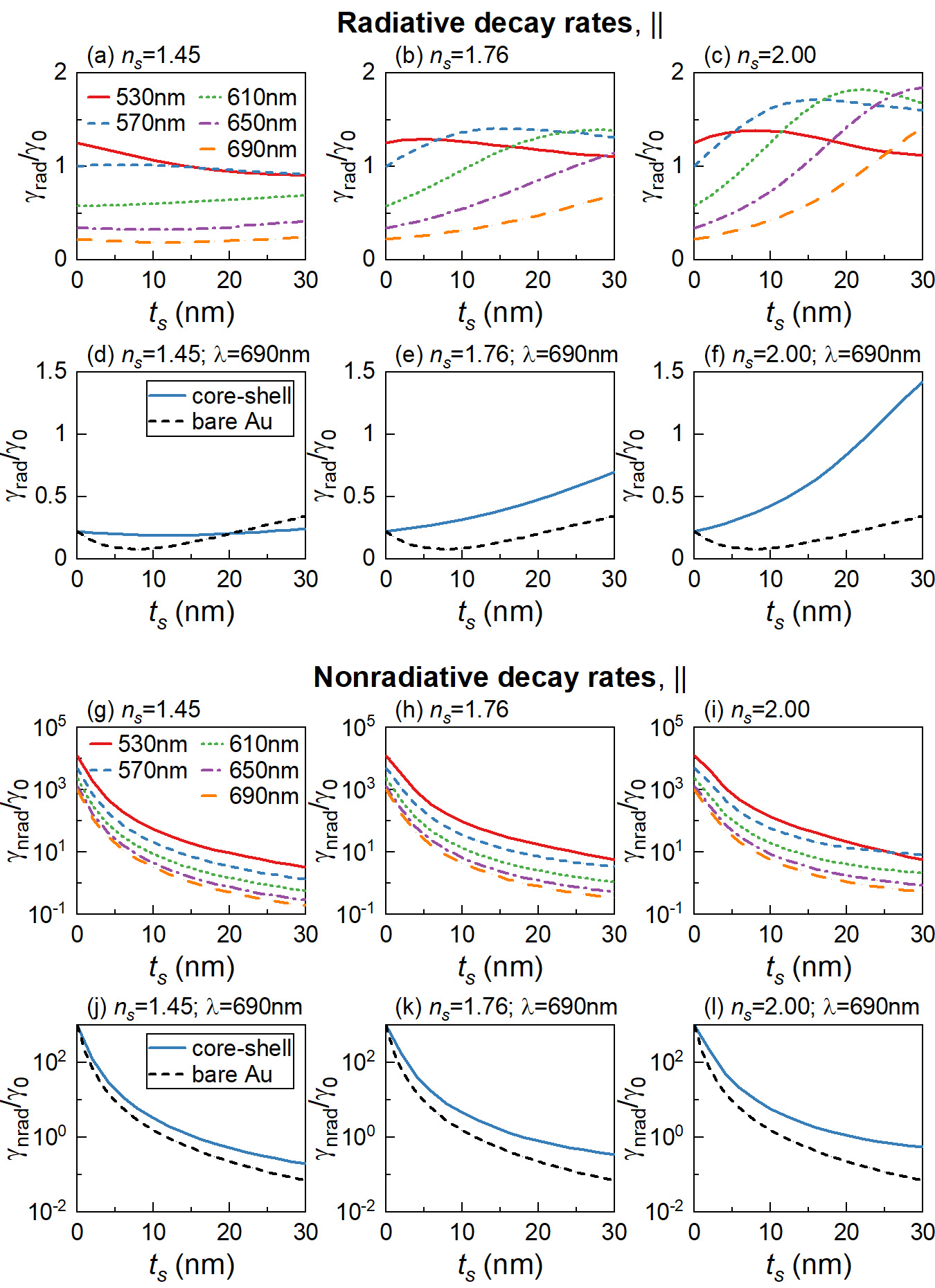}
	\caption{
	(a)-(c) Radiative and (g)-(i) nonradiative decay rates of parallel-oriented dipole emitter at the shell-medium interface for shell refractive indices $n_s=1.45$, 1.76 and 2.00. 
	Comparison of (d)-(f) radiative and (j)-(l) nonradiative decay rates at the shell-medium interface with those of Au sphere for $\lambda=690$~nm.
	Au core radius is $r_c=70$ nm in all cases.
	Influence of dielectric shell on parallel-oriented dipole emitter is similar to that of radially oriented emitter.}
	\label{fig:decparshll}
\end{figure}
%%%%%%%%%%%%%%

Figure~\ref{fig:decparshll} shows dipole decay rates at the shell-medium interface 
for a {\em parallel} oriented dipole emitter.  Figures~\ref{fig:decparshll}a--\ref{fig:decparshll}c and \ref{fig:decparshll}g--\ref{fig:decparshll}i show the radiative and nonradiative decay rates at the shell-medium interface for various shell refractive indices, and the comparisons between the core-shell structure and the Au sphere counterpart are illustrated in Figures~\ref{fig:decprpshll}d--\ref{fig:decprpshll}f and \ref{fig:decprpshll}j--\ref{fig:decprpshll}l, respectively. 
Both radiative and nonradiative decay rates at the shell-medium interface are generally larger compared to those of the Au sphere. 
Note that the absolute magnitudes of the dipole decay rates for the parallel-oriented emitter is much smaller than those for the radially oriented emitter, which is consistent with the results in Figure~4 and with literature~\cite{Moroz2005,Sun2016d,Zhang2019d}.

%The decrease of the dipole decay rates at the shell-medium interface could be beneficial to the fluorescence enhancement performance in terms of two aspects: 1) a high radiative decay rate obtained with a thick dielectric shell directly improves the fluorescence emission, and 2) although the nonradiative decay rate at the shell-medium interface is still stronger than that of the homogeneous Au sphere, the optimal emitter location for the core-shell structure could be much longer than that of the homogeneous Au sphere due to the decelerated decrease of the excited electric field (see Figure~\ref{fig:flds}). 
%Therefore, a comparable or even lower non-radiative decay rate might be achieved at the optimal condition of the core-shell nanoparticle, which further contributes to an enhanced fluorescence.

\subsection{Quantum yield}
%%%%%%%%%%%%%%%%%%%%%%%%%%

\begin{figure}[t!]
	\centering
	\includegraphics{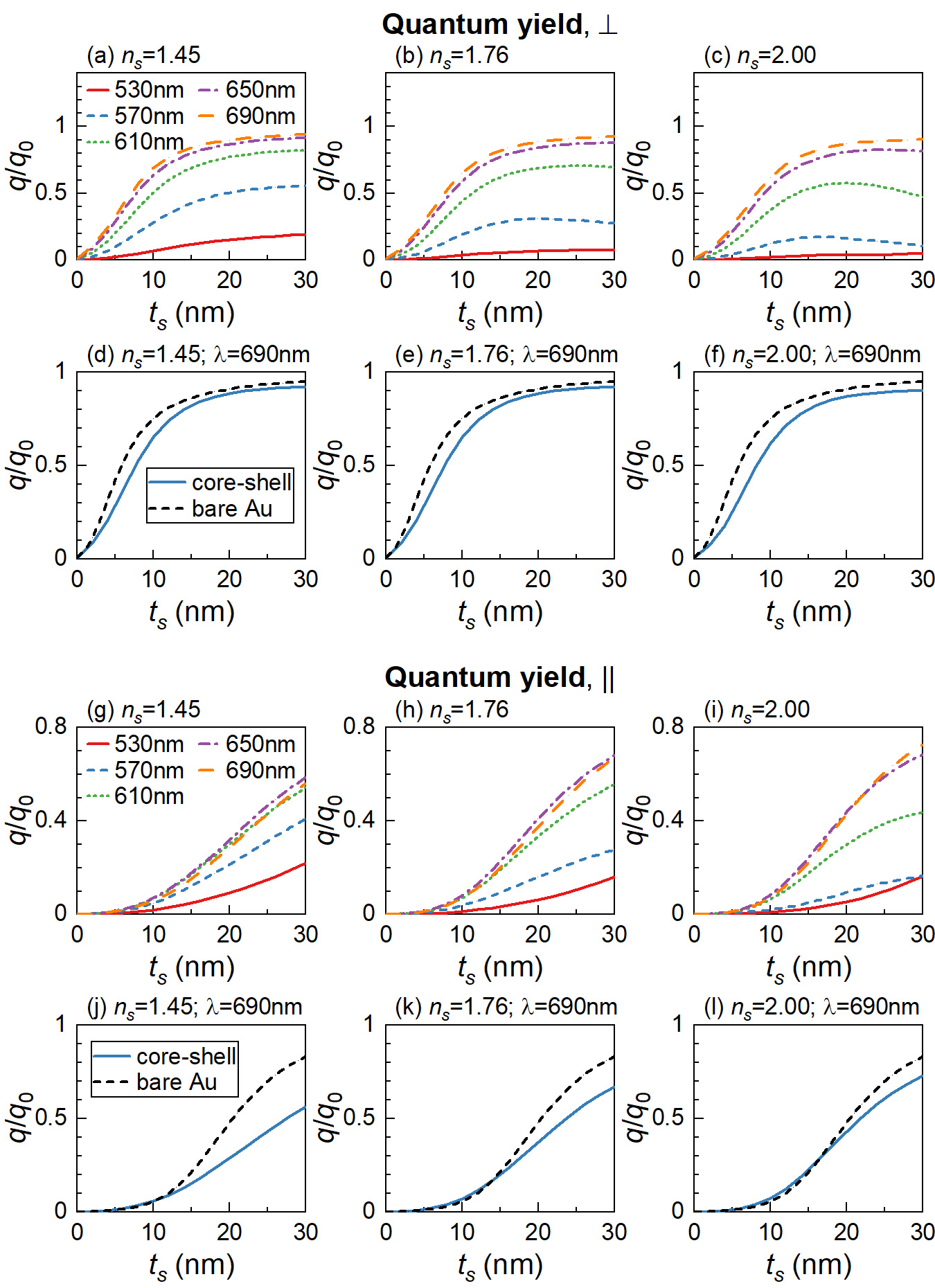}
	\caption{
	Quantum yields for 
	(a)-(c) perpendicular-oriented and 
	(g)-(i) parallel-oriented dipole emitters at the shell-medium interface for shell refractive indices $n_s=1.45$, 1.76 and 2.00. 	
	Comparison of quantum yields at the shell-medium interface with those around bare Au sphere for 
	(d)-(f) perpendicular-oriented and 
	(j)-(l) parallel-oriented emitters at $\lambda=690$~nm.
	Au core radius is $r_c=70$~nm in all cases.
	Core-shell particles exhibit similar quantum yields at the shell-medium interface compared to those around Au sphere at equivalent distances.}
	\label{fig:qy}
\end{figure}

Figures~\ref{fig:qy}a--\ref{fig:qy}c show the quantum yields for the perpendicular-oriented emitter at the shell-medium interface at various wavelengths. 
%%%%%%%%%%%%%%
%%%%%%%%%%%%%%
The quantum yields are generally larger at larger wavelength, because the radiative components are maintained at a comparable level regardless of the wavelength while the nonradiative decay rates are dramatically reduced at larger wavelength. 
In addition, the quantum yield increases as the shell thickness increases, because the radiative decay rates are relatively the same even for sufficiently thick shells, whereas the nonradiative decay rates decrease as the shell thickness increases. 
Figures~\ref{fig:qy}d--\ref{fig:qy}f compare the quantum yields at the shell-medium interface with those around Au sphere in the homogeneous medium at $\lambda = 690$~nm. 
The quantum yields at the shell-medium interface are slightly smaller than that of the bare Au particle at the equivalent distance, because both the radiative and nonradiative decay components are enhanced at the shell-medium interface.
Note that the quenching effect might still exist for a thin shell (i.e. $t_s<5$~nm), which should be avoided in the practical experiments~\cite{Aslan2007,Niu2018,Meng2018}. 
The conclusion also holds valid for the parallel-oriented emitter, which are shown in Figures~\ref{fig:qy}g--\ref{fig:qy}l. 
The profile of the quantum yield for the parallel-oriented emitter is different with that of the perpendicular-oriented emitter, which is attribute to the difference in the profile of radiative decay rates as shown in Figure~\ref{fig:decparshll}.
Note in passing that the comparison between core-shell particle and bare Au in Figure~\ref{fig:qy} is at the \textit{same} wavelength, and the quantum yield of the core-shell particle is slightly smaller than that of bare Au. Whereas the comparison in Figure 5 in the main text is at the corresponding optimal conditions of the two configurations, where the optimal wavelength of the core-shell particle is larger than that of bare Au, and the resultant quantum yield of the core-shell structure is larger than that of bare Au. 
%%%%%%%%%%%%%%
\begin{figure}
	\centering
	\includegraphics{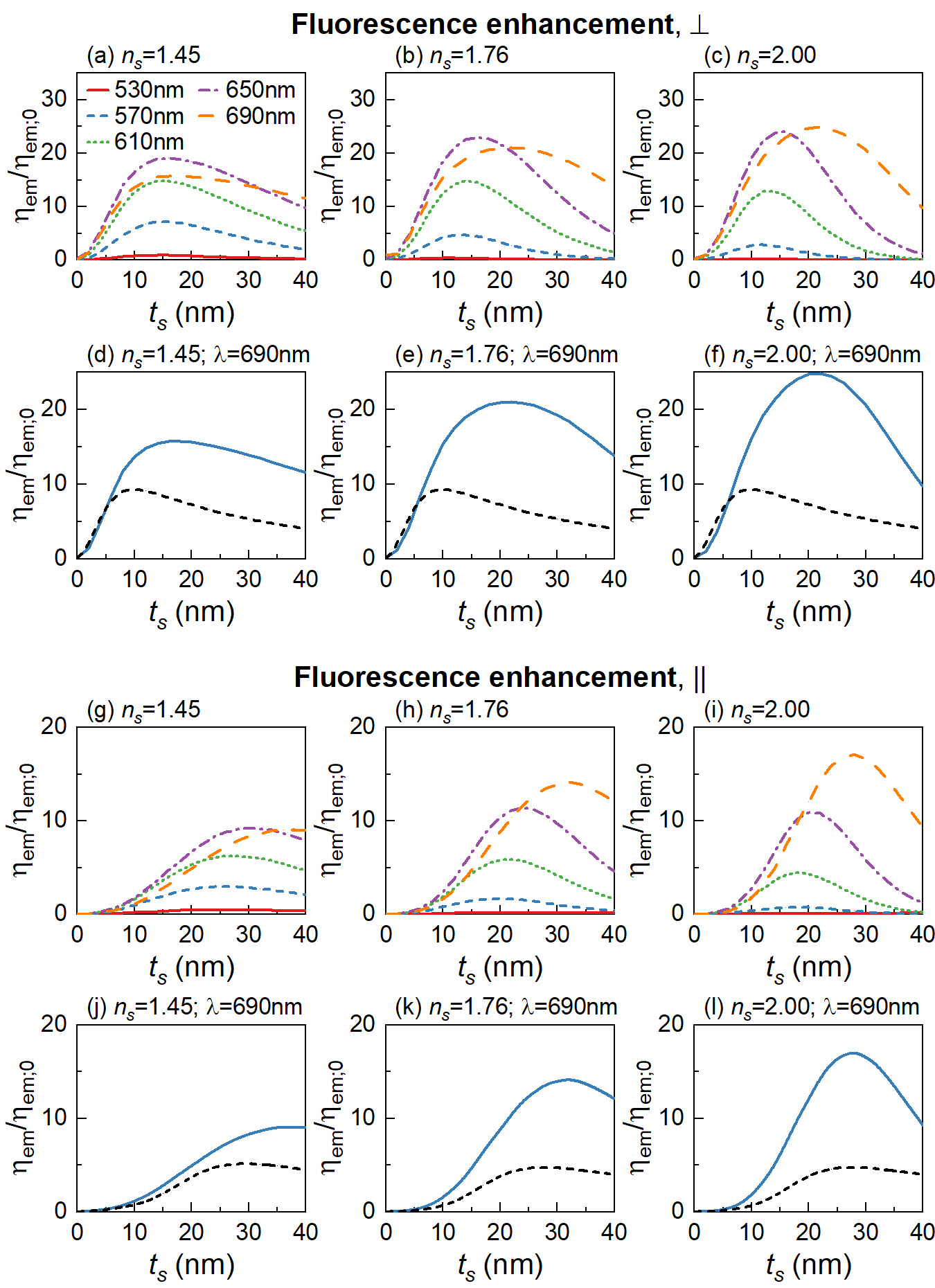}
    \caption{
    Fluorescence enhancement factors of (a)-(c) perpendicular-oriented and {\bf (g)-(i)} parallel-oriented emitters at the shell-medium interface for shell refractive indices $n_s=1.45$, 1.76 and 2.00. 
    Comparison of fluorescence enhancement factors at the shell-medium interface with those around Au sphere in homogeneous medium for (d)-(f) perpendicular-oriented and j-(l) parallel-oriented emitters at $\lambda=690$~nm.
    Au core radius is $r_c=70$ nm in all cases.
    Core-shell nanoparticle could generate much larger enhancement factor at the shell-medium interface than those of bare Au particle at the comparable metal-dye distance.}
    \label{fig:fl}
\end{figure}
%%%%%%%%%%%%%%

\subsection{Fluorescence enhancement}
%%%%%%%%%%%%%%%%%%%%%%%%%%

The ultimate fluorescence enhancement factor is obtained via multiplying the excitation rate and quantum yield. 
Figures~\ref{fig:fl}a--\ref{fig:fl}c illustrate the fluorescence enhancement factors for the perpendicular-oriented emitter at the shell-medium interface at various wavelengths and shell refractive indices.
In general, larger fluorescence enhancement factor is achieved at larger wavelength, due to the red-shift of the LSPR. 

Even larger fluorescence enhancement factor could be achieved by using a higher refractive index shell, which are attributed to its stronger capability in manipulating the decrease of excited electric fields and dipole decay rates. 
Figures~\ref{fig:fl}d--\ref{fig:fl}f compares the fluorescence enhancement factors at the shell-medium interface with those around Au sphere in the homogeneous medium at $\lambda=690$~nm. 
The core-shell nanoparticle could generate a much larger fluorescence enhancement factor at the shell-medium interface than that of the bare Au nanoparticle. 
On top of that, the maximum enhancement factors of the core-shell configurations appear at much longer distances than that of the bare Au particle, due to the decelerated decrease of electric field and dipole decay rates at the shell-medium interface. 
Similar results can also be obtained for the parallel-oriented emitter as shown in Figure~\ref{fig:fl}g--\ref{fig:fl}l. 
These findings are consistent with the optimizations results in Figures 2 and 3.

\section{Quasi-static formulas}
\lb{sc:GN}
%%%%%%%%%%%%%%%%%%%%%%%%%%%%%%%%%%
The Gersten and Nitzan (GN) quasi-static approximation for determining decay rates \cite{Gersten1981}, which makes use of particle multipolar polarizabilities, $\al_\ell$, has been known to provide a very good approximation of radiative and nonradiative decay rates in the case of small homogeneous particles \ct{Moroz2010}.
A dipolar polarizability of a core-shell particle with the account of dynamic depolarization and radiative correction \ct{Moroz2009} can be expressed as (cf. Eq.(23) in Ref.~\citenum{Chung2009}),
\bg
\al_{1}=\fr13\, r_2^3 
\fr{(\veps_2-\veps_h)[\veps_1q_1-\veps_2(q_1-1)]r_2^3
-
(\veps_1-\veps_2)[\veps_2 (q_2-1)-\veps_h q_2]r_1^3
}
{[\veps_1q_1-\veps_2(q_1-1)][\veps_2q_2-\veps_h (q_2-1)]r_2^3
-
(\veps_1-\veps_2)(\veps_2-\veps_h) q_2(q_2-1)r_1^3
},
\lb{mlwapol}
\eg
where
\bg
q_j:= \fr13 - \fr13\, x_j^2 -i \fr{2}{9}\,\, x_j^3,
\nn
\eg
$x_j=kr_j$, and the respective $r_1$ and $r_2$ are the radii of the inner 
and outer surfaces. In the case of an improved modified long wavelength approximation (IMLWA),
\bg
q_j:= \fr13 - \fr13\, x_j^2\left(1-\fr{2\bt}{5}\right) -i \fr{2}{9}\,\, x_j^3,
\lb{btpr}
\eg
where $\bt$ is a dimensionless fitting parameter to account for a spatially inhomogeneous polarization \ct{Moroz2009},
\bg
\vP \to (1+\bt k^2 r^2\sin^2\theta) \vP,
\eg
the choice of $\bt=1$ in eq~\ref{btpr} enabled rather precise description of the scattering properties of core-shell particles \cite{Chung2009}.

To accurately describe multipole polarizabilities, we have employed an effective sphere dielectric constant, $\bar \veps$:

\begin{equation}
    \bar \veps = \dfrac{1 + 2\al_1}{1 - \al_1} \ ,
\end{equation}
where $\al_1$ is determined by eq~\ref{mlwapol}.
Substitution of $\bar \veps$ into GN approximation~\cite{Gersten1981} didn't result in any significant improvement, as shown in Figure 6.

%%%%%%%%%%%%%%%%%%%%%%%%%%
\bibliography{references}
%%%%%%%%%%%%%%%%%%%%%%%%%%